\documentclass[journal,twocolumn]{IEEEtran}
\usepackage{graphicx}

\newcommand{\bstfile}{IEEE} 

\newcommand {\beq}{\begin{equation}}
\newcommand {\eeq}{\end{equation}}
\newcommand {\beqa}{\begin{eqnarray}}
\newcommand {\eeqa}{\end{eqnarray}}

\begin{document}
\title{A general theory of inhomogeneous broadening for nonlinear susceptibilities: the second
hyperpolarizability}
\author{Robert~J.~Kruhlak,
Mark~G.~Kuzyk
\thanks{ Robert J. Kruhlak is at the Department of Physics, University of Auckland, Auckland, NZ E-mail: r.kruhlak@auckland.ac.nz}%
\thanks{Mark G. Kuzyk is at the Department of
Physics and Astronomy, Washington State University Pullman, Washington 99164-2814 E-mail:
kuz@wsu.edu}}


\maketitle
\begin{abstract}
A general theory of inhomogeneous broadening is rarely applied to nonlinear spectroscopy in lieu of either a simple Lorentzian or Gaussian model. In this work, we generalize all the important third-order nonlinear susceptibility expressions
obtained with sum-over state quantum calculations to include Gaussian and stretched Gaussian distributions of Lorentzians.  This theory gives a better fit to subtle spectral
features - such as the shoulder of the electroabsorption peak, and is a more accurate tool for determining transition moments from spectroscopy experiments.
\end{abstract}
%

\section{Introduction}

\IEEEPARstart{S}{um}-over states quantum perturbation treatments of the $b^{th}$-order nonlinear susceptibility
tensor, $\xi_{ij...k}^{(b)}$, in the dipole approximation yields a sum of terms of the form:
 \beqa
 \lefteqn{\xi_{ij...k}^{(b)} \propto \sum_{n}^{\infty} \sum_{m}^{\infty}
  ...\sum_{l}^{\infty}} \nonumber \\
  &&\hspace{-0.3in} \frac {\left(\mu_{i}\right)_{gn} \left(\mu_{j}\right)_{nm}
  ... \left(\mu_{k}\right)_{lg}} {(\omega_{ng} - i \Gamma_{ng} -  \omega_1)
   (\omega_{mg}-i \Gamma_{mg} -  \omega_1 -  \omega_2) ...} ,
 \label{general} 
 \eeqa
where $\left( \mu_{i} \right)_{nm}$ is the $nm$-matrix element of the $i^{th}$ component of the
electric dipole operator, $\omega_{nm}$ the transition frequency (energy) between states $n$ and
$m$, $\omega_i$ the frequency of the $i^{th}$ optical field, and $\Gamma_{ng}$ the
phenomenological damping factor.  The numerator is a product of $b+1$ transition moments and the
denominator a product of $b$ energy terms.  For an isolated molecule, the damping factor
$\Gamma_{ng}$ is inversely proportional to the lifetime of state $n$ and is a measure of the width
of the peak in the spectrum of $\xi_{ij...k}^{(b)}$ associated with a transition between state $n$
and the ground state $g$.

In this work, we take into account the interaction of molecules with their surroundings using a stochastic model as we have reported for the first- and second-order susceptibility.\cite{kruhl08.01}  The technique is similar to Stoneham's approach for the linear susceptibility\cite{stoneh69.01} and Toussaere's calculation of the hyperpolarizability,\cite{touss93.01} who both used Guassian statistics.  In our work, we generalize the statistics to stretched exponentials, which are known to better model the interaction between a molecule and a system that is characterized by a distribution of sites such as a host polymer.\cite{ghebr95.04,ghebr97.01}

In our treatment of inhomogeneous broadening, each molecule in an ensemble is then viewed as having a
different transition frequency (energy), $\omega_{ng}$.  For the stretched Gaussian, the probability distribution is of the form
 \begin{equation}
 f_{ng}(\delta\omega_{ng})=\frac {1} {N(\gamma_{ng},\beta)}
   \exp \left[ - \left( \frac {\delta\omega_{ng} }
    {\gamma_{ng}} \right)^{2\beta} \right] ,
  \label{stretched}
 \end{equation}
where $\delta \omega_{ng}= \omega_{ng} - \bar{\omega}_{ng}$, $\bar{\omega}_{ng}$ is the mean value
of the transition frequency, $N(\gamma_{ng},\beta)$ the normalization factor, $\gamma_{ng}$ the
linewidth of the distribution and $\beta$ is the distribution of sites parameter. For most systems, $\beta=0$
for an infinitely broad distribution and $\beta =1$ for a single characteristic width. The susceptibility will then be of the form,
 \beqa
 \lefteqn{\left( \int_{- \bar{\omega}_{ng}}^\infty d(\delta \omega_{ng}) \int_{- \bar{\omega}_{mg}}^\infty
  d(\delta \omega_{mg}) ... \right)} \nonumber \\
 && \xi_{ij...k}^{(b)} (\omega_{ng}, \omega_{mg},...) f_{ng}(\delta\omega_{ng}) f_{mg}(\delta\omega_{mg}) ...
  \label{convolution} 
 \eeqa

Note that $N(\gamma,\beta)$ depends on $\beta$, and will be written as
 \beq
 N(\gamma_{ng},\beta) = \gamma_{ng} \sqrt{\pi} B(\beta),
 \eeq
 where,
 \beq
 B(\beta) =  \left[ \frac{1}{\gamma_{ng} \sqrt{\pi}} \int^{\infty}_{-\infty}
               \exp \left[ - \left( \frac {\delta\omega_{ng} }
    {\gamma_{ng}} \right)^{2\beta} \right] d(\delta\omega_{ng})\right]
 \label{Bbeta}
 \eeq
to remain compatible with previous inhomogeneous broadening representations that use Gaussian
statistics \cite{touss93.01, otomo95.01,kruhl99.01,kruhl99.02}. Note that such stretched Gaussian statistics are also observed in light scattering experiments\cite{kruhl99.02,kruhl97.01}.

In this paper, we derive the expressions for the most important   third-order susceptibilities.  The specific results for a Gaussian and stretched exponential are
presented and compared with data from a quadratic electroabsorption (third-order susceptibility)
experiment to illustrate the usefulness of the theory. All results are summarized in an extensive appendix.

\section{Third-Order Energy Denominators}

Similar to first- and second-order processes\cite{kruhl08.01}, the SOS Lorentzian energy
denominators for third-order processes are
 \beqa
 \lefteqn{D^L_{ln}(-\omega_{\sigma};\omega_1,\omega_2,\omega_3)  = {\mathbf{S}}_{1,2,3}\times}\nonumber\\
 &&\left\{ \left[ (\Omega_{lg} - \omega_{\sigma}) (\Omega_{lg}-\omega_3)(\Omega_{n g} - \omega_1) \right]^{-1}\right. +   \nonumber \\
 & & \:\: \left[ (\Omega_{lg} -\omega_3)(\Omega_{n g}^{*}+\omega_2)(\Omega_{n g}-\omega_1)
 \right]^{-1} +  \nonumber \\ &  & \:\: \left[ (\Omega_{lg}^{*}+
 \omega_{\sigma})(\Omega_{lg}^{*}+\omega_3)(\Omega_{n g}^{*}+\omega_1) \right]^{-1} + \nonumber
 \\  &  & \:\left. \: \left[(\Omega_{lg}^{*} + \omega_3)(\Omega_{n g} - \omega_2)(\Omega_{n
 g}^{*} + \omega_1) \right]^{-1} \right\}, \label{Dln}
 \eeqa
and
 \beqa
  \lefteqn{D^L_{lmn}(-\omega_{\sigma};\omega_1,\omega_2,\omega_3) = {\mathbf{S}}_{1,2,3}\times}\nonumber\\
  && \left\{
\left[ (\Omega_{lg} - \omega_{\sigma}) (\Omega_{mg}-\omega_1-\omega_2)(\Omega_{n g} - \omega_1)
\right]^{-1}
  \right. +   \nonumber \\
&  & \:\: \left[ (\Omega_{lg}^{*}+\omega_3)(\Omega_{mg}-\omega_1-\omega_2)(\Omega_{n g}-\omega_1)
\right]^{-1} +  \nonumber \\ &  & \:\: \left[ (\Omega_{lg}^{*}+
\omega_1)(\Omega_{mg}^{*}+\omega_1+\omega_2)(\Omega_{n g}-\omega_3) \right]^{-1} + \nonumber
\\ &  & \:\: \left.
\left[(\Omega_{lg}^{*}+\omega_1)(\Omega_{mg}^{*}+\omega_1+\omega_2)(\Omega_{n
g}^{*}+\omega_{\sigma}) \right]^{-1}\right\}.
 \label{Dlmn}
 \eeqa
In our notation, $D^L_{ln}(-\omega_{\sigma};\omega_1,\omega_2,\omega_3)$ represents interactions,
which involve only one-photon states, and $D^L_{lmn}(-\omega_{\sigma};\omega_1,\omega_2,\omega_3)$
represents interactions that involve both one- and two-photon states.

It is significantly more difficult to calculate the
third-order IB theory because of the triple product of Lorentzian terms in the denominator.
In order to transform $D^L_{ln}$, and/or $D^L_{lmn}$ for a specific
set of input and output frequencies, the number of excited states must be known prior to performing a
partial fraction expansion of each denominator term. For example if there are two
distinct excited states ($l$ and $n$), it is necessary to perform the following partial fraction
expansion,
 \beqa
  \frac{1}{(\Omega_{lg}-\omega)\Omega_{lg}(\Omega_{ng}-\omega)}&=&
  \frac{1}{\omega}\left[\frac{1}{(\Omega_{lg}-\omega)(\Omega_{ng}-\omega)} \right. \nonumber \\
                        && \left.- \frac{1}{\Omega_{lg}(\Omega_{ng}-\omega)}\right],
 \eeqa
to eliminate the product of the two $\Omega_{lg}$ terms. These type of expansions allow us to write
the nonlinear energy denominators in terms of $W^{(1)}_{\beta}(z)$ or when $\beta=1$ in terms of
complex error functions.

The first and second-order IB theory has been derived from the Lorentzian denominator terms by Kruhlak and Kuzyk,\cite{kruhl08.01} so we begin with the analogous derivation of the fundamental transformation for the third-order denominator to model all third-order processes.  As an example consider,
 \beq
 \frac{C_3}{(\omega'_{ng} -i\Gamma_{ng} -\omega)^3}.\label{edtoexamp}
 \eeq
 Equation (\ref{edtoexamp}) is multiplied by the stretched Gaussian function and is integrated with respect
to $\delta\omega_{ng}$,  the integration variable is changed to $t=(\omega^{\prime}_{ng} -
\omega_{ng})/\gamma_{ng}$, and $z$ replaces $(- \omega_{ng}+ i\Gamma_{ng}+\omega)/{\gamma_{ng}}$,
to get the following:
 \beqa
 \lefteqn{\int^{\infty}_{-\omega_{ng}} \frac{C_3}{(\omega'_{ng} -i\Gamma_{ng} -\omega)^3} f_{ng}
 (\omega'_{ng} -\omega_{ng}) d(\omega'_{ng} -\omega_{ng}) } \nonumber \\
 &=&\hspace{-0.09 in} \frac{C_3}{\gamma_{ng} \sqrt{\pi} B(\beta)}
  \int^{\infty}_{-\omega_{ng}} \frac{\exp({-(\frac{\omega'_{ng} -\omega_{ng}}{\gamma_{ng}})^{2\beta}})}
  {(\omega'_{ng} -i\Gamma_{ng} -\omega)^3} d(\omega'_{ng} -\omega_{ng}), \nonumber \\
 &=&\hspace{-0.09 in} \frac{C_3}{\gamma_{ng} \sqrt{\pi} B(\beta)}
 \int^{\infty}_{-\frac{\omega_{ng}}{\gamma_{ng}}}
 \frac{\gamma_{ng}\exp({-t^{2\beta}})}{(\omega'_{ng} -i\Gamma_{ng} -\omega)^3} dt, \nonumber \\
 &=&\hspace{-0.09 in} \frac{C_3}{\gamma_{ng}^3 \sqrt{\pi} B(\beta)}
 \int^{\infty}_{-\frac{\omega_{ng}}{\gamma_{ng}}}
 \frac{\exp({-t^{2\beta}})}{(t + \frac{\omega_{ng} -i\Gamma_{ng} -\omega}{\gamma_{ng}})^3} dt,
  \nonumber \\
 &=&\hspace{-0.09 in} \frac{-C_3}{\gamma_{ng}^3 \sqrt{\pi} B(\beta)}
 \int^{\infty}_{-\frac{\omega_{ng}}{\gamma_{ng}}} \frac{\exp({-t^{2\beta}})}{(z - t )^3} dt,
 \nonumber \\
 &=&\hspace{-0.09 in}\frac{i \sqrt{\pi} C_3}{\gamma_{ng} }\left[\frac{1}{\gamma^2_{ng}}W^{(3)}_{\beta}(z)\right]
 \label{C3transform}
 \eeqa
where \beq
  W(z)^{(3)}_{\beta} =\int^{\infty}_{-\frac{\omega_{ng}}{\gamma_{ng}}}
\frac{\exp({-t^{2\beta}})}{(z - t )^3} dt,
 \label{W3beta}
 \eeq
and $z = -(\omega_{ng} - i\Gamma_{ng} - \omega)/{\gamma_{ng}}$.

The integral in Eq. (\ref{W3beta}) looks similar to the complex error function\cite{abram72.01},
 \beq
  W(z) =\int^{\infty}_{-\frac{\omega_{ng}}{\gamma_{ng}}}
\frac{\exp({-t^{2}})}{(z - t )} dt,
 \eeq
 when $\beta=1$, except for the denominator. We use integration by parts to re-express the denominator to first-order in $(z-t)$.
With $T=\exp(-t^2)$, integrating by parts twice yields:
 \beqa
 \lefteqn{\int^{\infty}_{-\frac{\omega_{ng}}{\gamma_{ng}}} \frac{T}{(z - t )^3} dt
  = \int^{\infty}_{-\frac{\omega_{ng}}{\gamma_{ng}}} \frac{tT}{(z - t )^2} dt}\nonumber \\
&=& 2 \int^{\infty}_{-\frac{\omega_{ng}}{\gamma_{ng}}} \frac{t^2T}{(z - t )} dt -
\int^{\infty}_{-\frac{\omega_{ng}}{\gamma_{ng}}} \frac{T}{(z - t )} dt,
 \label{zt3}
 \eeqa
where certain terms vanish when the argument of the exponential is small ($\approx - 10^3$) at the lower limit.

Using $(z+t)= (z^2 - t^2)/(z-t)$ and
$\int_{-\infty}^{+\infty} t \exp(-t^2) dt = 0$ to recast Equation (\ref{zt3}) into a more convenient form, we get:
 \beqa
 \lefteqn{\int^{\infty}_{-\frac{\omega_{ng}}{\gamma_{ng}}} \frac{T}{(z - t )^3} dt
=(2z^2 -1) \int_{- \frac {\omega_{ng}} {\gamma_{ng}}}^{+\infty} \frac{T}{(z - t )}dt}  \nonumber
\\
& - & 2z\int^{\infty}_{-\frac{\omega_{ng}}{\gamma_{ng}}}T dt, \simeq (1- 2z^2) i \pi W(z) - 2z\sqrt{\pi}.
 \eeqa
So, the convolution of the cubic Lorentzian with the Gaussian distribution (with $\beta=1$) is,
 \beqa
 \lefteqn{\int^{\infty}_{-\infty} \frac{C_3}{(\omega'_{ng} -i\Gamma_{ng} -\omega)^3}
 f_{ng}(\omega'_{ng} -\omega_{ng}) d(\omega'_{ng} -\omega_{ng}) }\nonumber \\
 &=& \frac{-C_3}{\gamma_{ng}^3 \sqrt{\pi}} \left \{(1- 2z^2) i \pi W(z) - 2z\sqrt{\pi}\right \},
 \nonumber \\
 &=& \frac{ i  \sqrt{\pi} C_3}{\gamma_{ng}} \left \{\frac{ 2z^2 -1}{\gamma_{ng}^2}
 W(z) - \frac{2 i z}{ \sqrt{\pi}\gamma_{ng}^2}\right \},
 \eeqa
where $z=(- \omega_{ng}+ i\Gamma_{ng}+\omega)/{\gamma_{ng}}$.

Table \ref{tab:transform} summarizes the third-order fundamental energy denominators for the Lorentzian and IB theories with $\beta \leq 1$, and $\beta =1$, respectively.  These in conjunction with those derived in Ref. \cite{kruhl08.01} can be used to construct any IB energy denominator for any first-, second-,
and/or third-order process.
\begin{table}[!h]
\begin{center}
 \caption{{\label{tab:transform} Fundamental denominator
contributions to homogeneously broadened and inhomogeneously broadened electronic transitions} }
\begin{tabular}[t]{cc}   \hline
Model& Equation \\
\hline \\
L&  $\frac{C_3}{(\omega_{ng} \mp i\Gamma_{ng} \mp \omega)^3}$ \\
&\\
 IB ($\beta\leq1$) &$\frac{ i  \sqrt{\pi} C_3}{\gamma_{ng}}
 \left[ \frac{1}{\gamma^2_{ng}}
 W^{(3)}_{\beta}\left(\frac{-\omega_{ng} \pm i\Gamma_{ng} \pm \omega}{\gamma_{ng}}\right)
  \right ] $ \\
  &\\
 IB ($\beta = 1$)
 &$\frac{ i  \sqrt{\pi} C_3}{\gamma_{ng}}
 \left \{ \left(\frac{ 2(\omega_{ng} \mp i\Gamma_{ng} \mp \omega)^2 -
\gamma_{ng}^2}{\gamma_{ng}^4}\right) \times\right.$
\\
 &\\
  & $\left. W\left(\frac{-\omega_{ng} \pm i\Gamma_{ng} \pm
\omega}{\gamma_{ng}}\right) +  \frac{2 i (\omega_{ng} \mp i\Gamma_{ng} \mp
\omega)}{ \sqrt{\pi} \gamma_{ng}^3}\right \} $ \\
\\
  \hline
\end{tabular}
\end{center}
 \end{table}

\section{Third-Order Molecular Susceptibility}

\begin{figure}[!htb]
\begin{center}
\scalebox{0.4}{\includegraphics{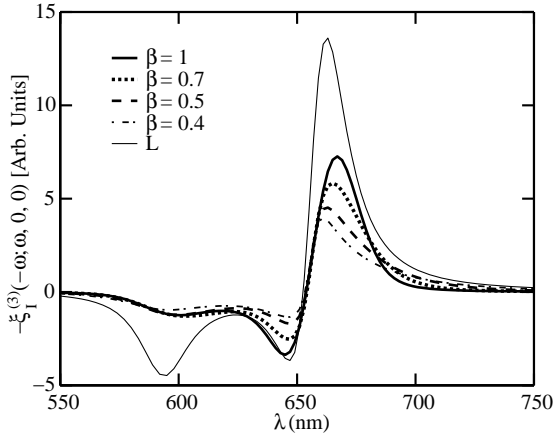}}
\end{center}
\caption{{Imaginary part of $\xi^{(3)}(-\omega; \omega, 0, 0)$ from the generalized IB theory for a one-photon excited state centered at 660 nm ($\Gamma_{1g}= 10 $ meV, and $\gamma_{1g}
= 40$ meV) and a two-photon state centered at 595 nm ($\Gamma_{2g}= 40 $ meV, and $\gamma_{2g} =
40$ meV).  Three values of $\beta$ are compared to the Lorentzian theory ($\Gamma_{1g} = 40$ meV,
and $\Gamma_{2g} = 40$ meV).  $\mu_{2g}/\mu_{1g} = 0.4$ for all the models. }}
 \normalsize \label{fig:chi3ibeta}
 \end{figure}
As an example of the third-order molecular susceptibility for homogeneous and inhomogeneous models,
we use Eq.'s (\ref{Dln}), (\ref{Dlmn}),
(\ref{Dllqeaibbetashort}),(\ref{Dlmlqeaibbetashort}),(\ref{Dllqeaibshort}) and
(\ref{Dlmlqeaibshort}), for the respective models, in
 \beqa
 \xi^{(3)}(-\omega;\omega,0,0) &=&
  \frac{1}{ \epsilon_0 3!}\frac{1}{ \hbar^3}
 \left\{ |\mu_{g1}|^2 |\mu_{12}|^2 D_{121}(-\omega; \omega, 0, 0)
    \right.
 \nonumber \\
 & &\hspace{0.3 in}\left. - |\mu_{g1}|^4 D_{11}(-\omega; \omega, 0, 0) \right\},
 \label{xi3l11p12pqea}
 \eeqa
to calculate the imaginary part of $\xi^{(3)}(-\omega; \omega,0,0)$ for a three-level system.
Figure \ref{fig:chi3ibeta} shows the imaginary part of the third-order susceptibility for a system
with one one-photon and one two-photon excited state for the quadratic electrooptic process. We see
a decrease in the magnitude of the third-order susceptibility from the Lorentzian theory to the IB
theory, which suggests that inhomogeneous broadening decreases the effective nonlinearity of the
material for this process. This trend continues as the distribution of sites becomes broader.
Additionally the significant reduction in the response near the center wavelength of the two-photon
state can be very dramatic when the contributions from the one- and two-photon states are of
similar strength. This occurs because they have opposite signs in Eq. (\ref{xi3l11p12pqea}).
Therefore it is important to use the IB theory for the guest-host system to model this particular region.

\section{Comparison of Theory to Quadratic Electroabsorption Experimental Results}

 \begin{figure}[!h]
 \begin{center}
 \scalebox{.65}{\includegraphics{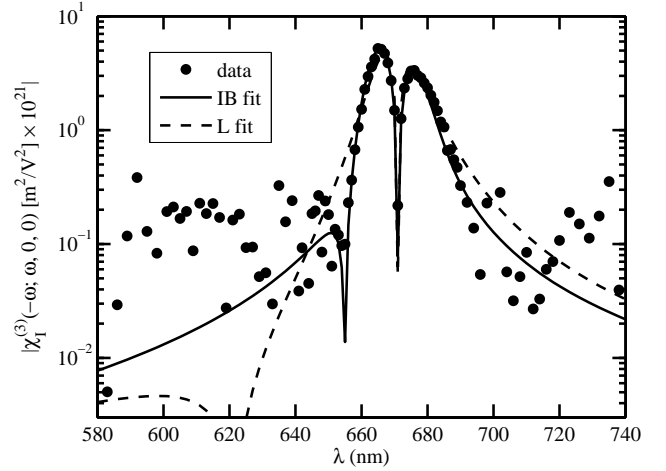}}
 \end{center}
 \caption{{Absolute value of the imaginary part of $\chi^{(3)}$ from a quadratic electroabsorption experiment on SiPc/PMMA and least-squares fits using Lorentzian
  and IB($\beta=1$) theories. The fit parameters are from ref. \cite{kruhl05.01}. }}
  \normalsize
  \label{fig:magimagchi3sipc}
  \end{figure}

We use quadratic electroabsorption spectra to test our models. Details of the experiment and the relationship between Eq.
(\ref{xi3l11p12pqea}) and $\chi^{(3)}(-\omega; \omega,0,0)$ can be found in the literature.
\cite{kruhl05.01} Figure \ref{fig:magimagchi3sipc} compares experimental values of the imaginary
part of $\chi^{(3)}$ for silicon phthalocyanine-methylmethacrylate in polymethylmethacrylate
(SiPc/PMMA), which were derived from quadratic electroabsorption experiments.  Also plotted are the Lorentzian (L)
and IB theories.  A log scale is used to highlight the qualitative and quantitative features of the two models. The fit parameters are from the literature\cite{kruhl05.01}. Like the linear absorption fits in Ref. \cite{kruhl08.01}, the IB model fits the
data better both quantitatively, roughly a factor of 6 smaller relative error, and qualitatively,
especially in the wings of the resonant signal. Neither model fits the data off-resonance because
of the large random error associated with the lock-in amplifier signal away from resonance. The
error bars cannot be plotted on a log scale because the error range includes negative values in the wings.

Nonlinear spectroscopy experiments aim to
determine zero frequency nonlinear susceptibilities by extrapolation, which can lead to large uncertainties depending on the quality of the dispersion models.  Indeed,
Canfield\cite{canfi05.01}, Vigil,\cite{vigil01.01} and Kruhlak \cite{kruhl05.01} have shown that it is often difficult to reconcile the transition moments as determined by independent means.  The wing region near resonance and the shape of the resonance peak may play an important role when using fitting to determine transition moments or for extrapolating to off-resonant values of $\chi^{(n)}$ from a data set with limited spectral range.

Differences between the IB model and the standard Lorentzian model can be used to determine the reliability of zero-frequency
susceptibilities and the uncertainty in transition moments.  More importantly, precise modeling aimed at understanding the dispersion of the nonlinear-optical response will need to take into account all possible broadening mechanisms.  Because IB theory takes into account the distribution of sites, it may well be the best way to model systems such as dye-doped polymers.

\section{Conclusion}

In conclusion, we have calculated the inhomogeneously broadened third-order nonlinear-optical
susceptibilities for a Gaussian and stretched Gaussian distribution of Lorentzians.  The results
are applied to the quadratic electro-absorption spectrum of SiPc/PMMA and we find that the
Lorentzian fit alone does not fit the data at the wings.  The IB theory, however, fits the data
over a broader wavelength range and shows that the distribution of sites is nearly Gaussian,
implying that interactions between the polymer and dopant are small.

Since broadening of the nonlinear susceptibility is shown to have an important affect the
dispersion, the determination of excited state properties of molecules from spectroscopy requires
that such a theory be used.  So, IB theory using stretched Gaussian statistics may become an
important tool for interpreting nonlinear-optical spectroscopy measurements.

 \section*{Acknowledgments} We thank the National Science
Foundation (ECS-0354736) and Wright Patterson Air Force Base for generously supporting this work.

\bibliographystyle{\bstfile}

\appendices

\section{Compact notation}

The energy denominators $D$ for the higher-order susceptibilities are complex combinations of
$W^{(x)}_{\beta}(z)$ or $W(z)$. We have developed a more compact notation than previously
used\cite{kruhl05.01}. For $\beta \leq 1$, we have added a subscript to $\beta$ to indicate the
excited state involved in the process and a $^*$ on the power of $W$ to indicate a complex
conjugate of the complex argument $\Omega$. Similarly for $\beta=1$,  the subscript on $W$
indicates the excited state and the superscript $^*$ on $W$ indicates the complex conjugate of
$\Omega$. This allows us to use a simple frequency argument of $\pm \omega$ that significantly
improves the readability of the equations in the extensive appendices that follow. An example of
the compact notation is given below: \beq W^{(1)}_{\beta}\left(\frac{-(\Omega^*_{2g} +
\omega_3)}{\gamma_{2g}}\right) \rightarrow W^{(1)^*}_{\beta_2}(\omega_3) . \eeq Since all arguments
are all of the  form $\frac{-(\Omega^*_{ng} \pm \omega_i)}{\gamma_{ng}}$ or $\frac{-(\Omega_{ng}
\mp \omega_i)}{\gamma_{ng}}$ and all transitions are from/to the ground state ($g$), this form
describes all inhomogeneous broadening terms in this paper. Table \ref{tab:compact} summarizes the
compact notation.

\begin{table}[!h]
\begin{center}
 \caption{{\label{tab:compact} Compact form of $W^{(x)}_{\beta}(z)$ and $W(z)$ up to third order($x =\{1,2,3\}$).} }
\begin{tabular}[!h]{ccc}
$\beta$ & Ref. \cite{kruhl05.01}& Compact Form \\\hline
 $\leq 1$& $
  W^{(x)}_{\beta}\left(\frac{-(\Omega_{ng} \mp
\omega_i)}{\gamma_{ng}}\right) $ & $W^{(x)}_{\beta_n}(\mp\omega_i) $\\
& $W^{(x)}_{\beta}\left(\frac{-(\Omega^*_{ng} \pm \omega_i)}{\gamma_{ng}}\right)$ &
$W^{(x)^*}_{\beta_n}(\pm\omega_i)$\\
  &&\\
 $1$& $W\left(\frac{-(\Omega_{lg} \mp \omega_i)}{\gamma_{lg}}\right)$ & $W_l(\mp\omega_i)$ \\
&$W\left(\frac{-(\Omega^*_{lg} \pm \omega_i)}{\gamma_{lg}}\right)$ & $W^*_l(\pm\omega_i)$\\
  \hline \\
\end{tabular}
\end{center}
 \end{table}

\onecolumn
\section{Energy Denominators for Selected Processes}

\subsection{Third-Harmonic Generation}
\subsubsection{$\beta \leq 1$}
 \beqa
D_{ll}^{IB}(-3\omega;\omega ,\omega,\omega)  &=&
\frac{\sqrt{\pi}}{\omega \gamma_{lg}}
 \left\{ \frac{i}{4 \omega}
 \left[ W^{(1)}_{\beta_l}(- 3\omega)
 + W^{(1)^{*}}_{\beta_{l}}(3\omega)
 - W^{(1)}_{\beta_l}(-\omega)
 - W^{(1)^{*}}_{\beta_l}(\omega) \right ] \right.\nonumber \\
 & & \hspace{0.43in} \left. + \frac{ \Gamma_{lg}}{2(\omega +i\Gamma_{lg})}
 \left [\frac{1}{\gamma_{lg}}
 W^{(2)^{*}}_{\beta_l}(\omega)
  - \frac{1}{\gamma_{lg}}
 W^{(2)}_{\beta_l}( - \omega)
   \right ] \hspace{0.1 in}\right\}
   \label{Dllthgibbetashort}
 \eeqa
 \beqa
D_{ln}^{IB}(-3\omega;\omega,\omega,\omega)
 & = &
 \frac{-\pi}{2\gamma_{lg}\gamma_{ng}}
 \left \{\frac{1}{\omega}
 \left[ W^{(1)}_{\beta_n}( - \omega)
 \left \{W^{(1)}_{\beta_l}( - 3\omega)
 - W^{(1)}_{\beta_l}(-\omega) \right \}
  \right.
+ W^{(1)^{*}}_{\beta_n}(\omega)
 \left \{W^{(1)^{*}}_{\beta_l}(\omega)
 - W^{(1)^{*}}_{\beta_l}( 3\omega) \right \} \right]
 \nonumber \\
 & &  \hspace{0.53 in} + \frac{1}{(\omega + i\Gamma_{ng})}
 \left[ \left \{ W^{(1)}_{\beta_l}(-\omega)
 +  W^{(1)^{*}}_{\beta_l}(\omega) \right \}
  \left.
 \left \{ W^{(1)}_{\beta_n}(-\omega)
 - W^{(1)^{*}}_{\beta_n}(\omega) \right \} \right ]
  \hspace{0.125 in}\right \}
   \label{Dlnthgibbetashort}
 \eeqa
 \beqa
 D_{lml}^{IB}(-3\omega;\omega,\omega,\omega)  &=& \frac{-\pi}{2\gamma_{lg}\gamma_{mg}}
 \left \{ \frac{1}{\omega}
 \left[ W^{(1)}_{\beta_m}(-2\omega) \left \{W^{(1)}_{\beta_l}(-3\omega)- W^{(1)}_{\beta_l}(-\omega) \right \}
  + W^{(1)^{*}}_{\beta_m}(2\omega)
 \left \{W^{(1)^{*}}_{\beta_l}(\omega) - W^{(1)^{*}}_{\beta_l}(3\omega) \right \} \right] \right.
 \nonumber \\
 & &  \hspace{0.53 in} \left. + \frac{1}{(\omega + i\Gamma_{lg})}
 \left[ \left \{ W^{(1)}_{\beta_m}(-2\omega)
 +  W^{(1)^{*}}_{\beta_m}(2\omega) \right \}
 \left \{ W^{(1)}_{\beta_l}(-\omega)
 - W^{(1)^{*}}_{\beta_l}(\omega) \right \} \right ]
  \hspace{0.25 in} \right \}
   \label{Dlmlthgibbetashort}
 \eeqa
 \beqa
  D_{lmn}^{IB}(-3\omega;\omega,\omega,\omega)  &=&
  \frac{-i\pi^{3/2}}{\gamma_{lg}\gamma_{mg}\gamma_{ng}}
 \left \{
 W^{(1)}_{\beta_n}(-\omega)W^{(1)}_{\beta_m}(-2\omega)
  \left [ W^{(1)}_{\beta_l}(-3\omega)+ W^{(1)^{*}}_{\beta_l}(\omega) \right ]
  \right.   \nonumber \\
 & & \hspace{0.7 in}\left. +
 W^{(1)^{*}}_{\beta_l}(\omega)W^{(1)^{*}}_{\beta_m}(2\omega)
 \left [ W^{(1)}_{\beta_n}(-\omega)+ W^{(1)^{*}}_{\beta_n}(3\omega) \right] \right\}
 \label{Dlmnthgibbetashort}
 \eeqa
\subsubsection{$\beta = 1$}
 \beqa
D_{ll}^{IB}(-3\omega;\omega ,\omega,\omega) & =&
\frac{\sqrt{\pi}}{\omega \gamma_{lg}}
\left\{ \frac{i}{4 \omega}
 \left[ W_{l}( - 3\omega) + W^{*}_{l}( 3\omega)- W_{l}( - \omega) - W^{*}_{l}( \omega) \right ] \right.\nonumber \\
 & & \hspace{0.53 in}\left.
 + \frac{ \Gamma_{lg}}{(\omega+i\Gamma_{lg})} 
 \left [\frac{(\Omega_{lg} - \omega)}{\gamma_{lg}^2}
 W_{l}( - \omega) - \frac{(\Omega^{*}_{lg} + \omega)}{\gamma_{lg}^2} W^{*}_{l}( \omega)
  \right ] \right\}
  \label{Dllthgibshort}  
 \eeqa
 \beqa
D_{ln}^{IB}(-3\omega;\omega,\omega,\omega)&=&
\frac{-\pi}{2\gamma_{lg}\gamma_{ng}}
 \left \{\frac{1}{\omega} \left[ W_{n}( - \omega) \left \{W_{l}( - 3\omega)- W_{l}( - \omega) \right \}
   + W^{*}_{n}( \omega) \left \{W^{*}_{l}( \omega) - W^{*}_{l}( 3\omega) \right \}
   \right] \right.
 \nonumber \\
 & &  \hspace{0.6 in} \left. + \frac{1}{(\omega + i\Gamma_{ng})}
 \left[ \left \{ W_{l}( - \omega) +  W^{*}_{l}( \omega) \right \}
 \left \{ W_{n}( - \omega) - W^{*}_{n}(\omega) \right \} \right ]
  \hspace{0.125 in}\right \}
   \label{Dlnthgibshort}
 \eeqa
 \beqa
 D_{lml}^{IB}(-3\omega;\omega,\omega,\omega)  &=& \frac{-\pi}{2\gamma_{lg}\gamma_{mg}}
 \left \{\frac{1}{\omega} \left[ W_{m}(- 2\omega)\left \{W_{l}( - 3\omega) - W_{l}( - \omega) \right \}
 + W^{*}_{m}( 2\omega)
 \left \{W^{*}_{l}( \omega)- W^{*}_{l}( 3\omega) \right \} \right]
 \right.
 \nonumber \\
 & & \hspace{0.6 in} \left. + \frac{1}{(\omega + i\Gamma_{lg})}
 \left[ \left \{ W_{m}(- 2\omega) +  W^{*}_{m}( 2\omega) \right \}
 \left \{ W_{l}( - \omega) - W^{*}_{l}( \omega) \right \} \right ]
  \hspace{0.025 in} \right \}
   \label{Dlmlthgibshort}
 \eeqa
 \beq
  D_{lmn}^{IB}(-3\omega;\omega,\omega,\omega)=
  \frac{-i\pi^{3/2}}{\gamma_{lg}\gamma_{mg}\gamma_{ng}}
 \left \{
 W_{n}(-\omega)W_{m}(- 2\omega)\left [W_{l}( - 3\omega)+ W^{*}_{l}( \omega) \right
 ] +
 W^{*}_{l}( \omega)
 W^{*}_{m}( 2\omega)
 \left [ W_{n}(-\omega) + W^{*}_{n}(3\omega) \right] \right\}
 \label{Dlmnthgibshort}
 \eeq
\subsection{Quadratic Electrooptic Effect}
\subsubsection{$\beta \leq 1$}
 \beqa
 D_{ll}^{IB}(-\omega;\omega,0,0) \hspace{-0.09 in}& =&
  \frac{2i\sqrt{\pi}}{\gamma_{lg}}
  \left\{ \frac{1}{\omega \gamma_{lg}}\left[
    W^{(2)^{*}}_{\beta_l}(\omega)
     - W^{(2)}_{\beta_l}(-\omega)
     \right]
     + \frac{ i \Gamma_{lg}}{{\omega}(\omega + 2i\Gamma_{lg})\gamma_{lg}}
 \left[ W^{(2)}_{\beta_l}(0)
  - W^{(2)^{*}}_{\beta_l}(0)
  \right]
  \right.
  \nonumber \\ & & \hspace{0.5 in} \left.
 + \frac{(1 + 2\frac{\Gamma_{lg}^2}{\omega^2})}{(\omega + 2i\Gamma_{lg})^2} \left[
 W^{(1)}_{\beta_l}(-\omega) + W^{(1)^{*}}_{\beta_l}(\omega)
 -  W^{(1)^{*}}_{\beta_l}(0)- W^{(1)}_{\beta_l}(0)\right]
 \right\}
  \label{Dllqeaibbetashort}
 \eeqa
 \beqa
 D_{ln}^{IB}(-\omega;\omega,0,0)  &=& \frac{-\pi}{\gamma_{lg}\gamma_{ng}}
 \left\{ W^{(1)}_{\beta_n}(0)
 \left[\frac{-1}{\gamma_{lg}}
  W^{(2)}_{\beta_l}(-\omega)
  \right] + W^{(1)^{*}}_{\beta_n}(0)
 \left[\frac{-1}{\gamma_{lg}}
 W^{(2)^{*}}_{\beta_l}(\omega) \right] \right.
 \nonumber
  \\
 & &  + \frac{1}{\omega}
 \left[ \left\{ W^{(1)^{*}}_{\beta_l}(0) - W^{(1)^{*}}_{\beta_l}(\omega)\right\}
  \left\{ W^{(1)^{*}}_{\beta_n}(\omega) + W^{(1)^{*}}_{\beta_n}(0)\right\} + \right. \nonumber \\
  && \hspace{0.4in}\left.
 \left\{ W^{(1)}_{\beta_l}(-\omega) - W^{(1)}_{\beta_l}(0)\right\}
  \left\{ W^{(1)}_{\beta_n}(-\omega) + W^{(1)}_{\beta_n}(0)\right\} \right]
  \nonumber \\
  & & + \frac{1}{2i\Gamma_{ng}}
  \left[ \left\{ W^{(1)}_{\beta_l}(-\omega)
  + W^{(1)^{*}}_{\beta_l}(\omega) \right\}
  \left\{ W^{(1)}_{\beta_n}(0) - W^{(1)^{*}}_{\beta_n}(0)\right\}\right]
  \nonumber \\
  & & \left. + \frac{1}{(\omega+2i\Gamma_{ng})}\left[
  \left\{W^{(1)}_{\beta_n}(0) - W^{(1)^{*}}_{\beta_n}(0) +  W^{(1)}_{\beta_n}(-\omega)- W^{(1)^{*}}_{\beta_n}(\omega) \right\}
  \left\{ W^{(1)}_{\beta_l}(0) + W^{(1)^{*}}_{\beta_l}(0) \right\}\right]
  \right \}
 \label{Dlnqeaibbetashort}
 \eeqa
 \beqa
 D_{lml}^{IB}(-\omega;\omega,0,0) & =&  \frac{-\pi}{\gamma_{lg}\gamma_{mg}}
 \left\{ W^{(1)}_{\beta_m}(-\omega) \left[\frac{-1}{\gamma_{lg}} W^{(2)}_{\beta_l}(-\omega) \right]
 + \frac{2(1 + i\frac{\Gamma_{lg}}{\omega})}{(\omega +2i\Gamma_{lg})}
 \left[W^{(1)}_{\beta_m}(-\omega)  W^{(1)}_{\beta_l}(-\omega)
 - W^{(1)^{*}}_{\beta_m}(\omega)W^{(1)^{*}}_{\beta_l}(\omega)
 \right] \right.
 \nonumber \\
 & & + W^{(1)^{*}}_{\beta_m}(\omega)
 \left[\frac{-1}{\gamma_{lg}}
 W^{(2)^{*}}_{\beta_l}(\omega) \right] + \frac{(1 + i\frac{\omega}{2\Gamma_{lg}})}{\omega}
 \left[ W^{(1)^{*}}_{\beta_m}(\omega) W^{(1)^{*}}_{\beta_l}(0)
 - W^{(1)}_{\beta_m}(-\omega) W^{(1)}_{\beta_l}(0)\right]
 \nonumber \\
 & & + \frac{2(1 - i\frac{\omega}{4\Gamma_{lg}})}{(\omega +2i\Gamma_{lg})}
 \left[ W^{(1)^{*}}_{\beta_m}(\omega) W^{(1)}_{\beta_l}(0) - W^{(1)}_{\beta_m}(-\omega) W^{(1)^{*}}_{\beta_l}(0) \right]
 \nonumber \\
 & & + \frac{2i\frac{\Gamma_{lg}}{\omega}}{(\omega + 2i\Gamma_{lg})}
 \left[W^{(1)^{*}}_{\beta_m}(0) W^{(1)^{*}}_{\beta_l}(0)
 - W^{(1)}_{\beta_m}(0) W^{(1)}_{\beta_l}(0)\right] \nonumber \\
 & &   + \frac{1}{\omega}
 \left[W^{(1)}_{\beta_m}(0) W^{(1)}_{\beta_l}(-\omega)
 - W^{(1)^{*}}_{\beta_m}(0) W^{(1)^{*}}_{\beta_l}(\omega) \right]
 \nonumber \\
 && \left.
  + \frac{1}{(\omega+2i\Gamma_{lg})}
 \left[ W^{(1)^{*}}_{\beta_m}(0) W^{(1)}_{\beta_l}(-\omega)
 - W^{(1)}_{\beta_m}(0) W^{(1)^{*}}_{\beta_l}(\omega)\right]  \right \}
 \label{Dlmlqeaibbetashort}
 \eeqa
 \beqa
 D_{lmn}^{IB}(-\omega;\omega,0,0) &=&
 \frac{-i\pi^{3/2}}{\gamma_{lg}\gamma_{mg}\gamma_{ng}}
 \left\{ W^{(1)}_{\beta_l}(-\omega)
 \left[W^{(1)}_{\beta_m}(-\omega) \left\{ W^{(1)}_{\beta_n}(-\omega) + W^{(1)}_{\beta_n}(0) \right\}
 + W^{(1)}_{\beta_m}(0) W^{(1)}_{\beta_n}(0)\right] \right.
 \nonumber \\
 & & + W^{(1)^{*}}_{\beta_l}(0)
 \left[ W^{(1)}_{\beta_m}(-\omega)
 \left\{W^{(1)}_{\beta_n}(-\omega) + W^{(1)}_{\beta_n}(0) \right\}  \right.
 \nonumber \\
 & & \hspace{0.7 in} \left. + W^{(1)^{*}}_{\beta_m}(\omega)
 \left\{ W^{(1)}_{\beta_n}(0) + W^{(1)^{*}}_{\beta_n}(\omega) \right\} +
 W^{(1)^{*}}_{\beta_m}(0)
 \left\{W^{(1)}_{\beta_n}(-\omega) + W^{(1)^{*}}_{\beta_n}(\omega) \right\} \right]
 \nonumber \\
 & & \left.+ W^{(1)^{*}}_{\beta_l}(\omega)\left[ W^{(1)^{*}}_{\beta_m}(\omega)
 \left\{W^{(1)}_{\beta_n}(0) + W^{(1)^{*}}_{\beta_n}(\omega) \right\} +
  W^{(1)}_{\beta_m}(0)
 W^{(1)}_{\beta_n}(0) \right]  \right \}
  \label{Dlmnqeaibbetashort}
 \eeqa
\subsubsection{$\beta = 1$}
 \beqa
 \hspace{-0.1in} D_{ll}^{IB}(-\omega;\omega,0,0) & =&
  \frac{2i\sqrt{\pi}}{\gamma_{lg}}
  \left\{  \frac{(1 + 2\frac{\Gamma_{lg}^2}{\omega^2})}{(\omega + 2i\Gamma_{lg})^2} \left[
 W_{l}( - \omega) + W^{*}_{l}(
  \omega) -  W^{*}_{l}(0)
  - W_{l}(0)
  \right] \right.
  \nonumber \\
& &\left.+ \frac{2}{\omega \gamma_{lg}^2}\left[(\Omega_{lg} -
\omega)
    W_{l}( - \omega)   -  (\Omega^{*}_{lg} + \omega)W^{*}_{l}( \omega) \right]
    +  \frac{ 2i \Gamma_{lg}}{{\omega}(\omega + 2i\Gamma_{lg})\gamma_{lg}^2}
 \left[\Omega^{*}_{lg} W^{*}_{l}(0)
  - \Omega_{lg} W_{l}(0)
  \right]
  \right\}
  \label{Dllqeaibshort}
 \eeqa
 \beqa
 D_{ln}^{IB}(-\omega;\omega,0,0)  &=&  \frac{-\pi}{\gamma_{lg}\gamma_{ng}}
  \left\{ W_{n}(0)
  \left[\frac{2(\Omega_{lg} - \omega)}{\gamma_{lg}^2}W_{l}( - \omega) +
  \frac{2i}{\sqrt{\pi}\gamma_{lg}}
 \right] + W^*_{n}(0)
 \left[\frac{2(\Omega^{*}_{lg} + \omega)}{\gamma_{lg}^2} W^{*}_{l}( \omega) + \frac{2i}{\sqrt{\pi}\gamma_{lg}}
 \right] \right.
 \nonumber
  \\
 & &  + \frac{1}{\omega}
 \left[ \left\{ W^{*}_{l}(0)
 - W^{*}_{l}( \omega)\right\}
  \left\{ W^{*}_{n}(\omega)
  + W^*_{n}(0)\right\}  +
 \left\{ W_{l}(- \omega)- W_{l}(0)\right\}
  \left\{ W_{n}(-\omega)+ W_{n}(0)\right\} \right]
  \nonumber \\
  & & + \frac{1}{2i\Gamma_{ng}}
  \left[ \left\{ W_{l}(- \omega)+ W^{*}_{l}( \omega) \right\}
  \left\{ W_{n}(0)- W^*_{n}(0)\right\}\right]
  \nonumber \\
  & & \left. +\frac{1}{(\omega+2i\Gamma_{ng})}\left[
  \left\{W_{n}(0) - W^*_{n}(0) + W_{n}(-\omega) - W^{*}_{n}(\omega) \right\}
  \left\{ W_{l}(0) + W^{*}_{l}(0) \right\} \right]  \right \}
 \label{Dlnqeaibshort}
 \eeqa
 \beqa
 D_{lml}^{IB}(-\omega;\omega,0,0)  &=&  \frac{-\pi}{\gamma_{lg}\gamma_{mg}}
 \left\{ W_{m}(- \omega)
 \left[\frac{2(\Omega_{lg} - \omega)}{\gamma_{lg}^2} W_{l}( - \omega) + \frac{2i}{\sqrt{\pi}\gamma_{lg}}\right]
  + W^{*}_{m}( \omega)
 \left[\frac{2(\Omega^{*}_{lg} + \omega)}{\gamma_{lg}^2} W^{*}_{l}( \omega) + \frac{2i}{\sqrt{\pi}\gamma_{lg}}
 \right] \right.
 \nonumber \\
 & &  + \frac{2(1 + i\frac{\Gamma_{lg}}{\omega})}{(\omega +2i\Gamma_{lg})}
 \left[W_{m}(- \omega) W_{l}( - \omega)- W^*_{m}(\omega) W^{*}_{l}( \omega) \right]
  + \frac{(1 + i\frac{\omega}{2\Gamma_{lg}})}{\omega}
 \left[ W^{*}_{m}( \omega)
 W^*_{l}(0)
 - W_{m}(- \omega)
 W_{l}(0)\right]
 \nonumber \\
 & & + \frac{2(1 - i\frac{\omega}{4\Gamma_{lg}})}{(\omega +2i\Gamma_{lg})}
 \left[ W^{*}_{m}( \omega)
 W_{l}(0)
 - W_{m}(- \omega)
 W^{*}_{l}(0) \right]
  + \frac{2i\frac{\Gamma_{lg}}{\omega}}{(\omega + 2i\Gamma_{lg})}
 \left[W^*_{m}(0)
 W^{*}_{l}(0)
 - W_{m}(0)
 W_{l}(0)\right] \nonumber \\
 & &  + \frac{1}{\omega}\left[W_{m}(0)
 W_{l}( - \omega)
 - W^*_{m}(0)
 W^{*}_{l}( \omega) \right]
 \nonumber \\
 & & + \left.\frac{1}{(\omega+2i\Gamma_{lg})}
 \left[ W^*_{m}(0)
 W_{l}( - \omega) -
 W_{m}(0)
 W^{*}_{l}( \omega)\right]  \right \}
 \label{Dlmlqeaibshort}
 \eeqa
 \beqa
D_{lmn}^{IB}(-\omega;\omega,0,0) &=&
 \frac{-i\pi^{3/2}}{\gamma_{lg}\gamma_{mg}\gamma_{ng}}
 \left\{ W_{l}( - \omega)
 \left[W_{m}(- \omega)
 \left\{ W_{n}(-\omega) + W_{n}(0) \right\}
 + W_{m}(0)W_{n}(0)\right] \right.
 \nonumber \\
 & & + W^{*}_{l}(0)
 \left[ W_{m}(- \omega)\left\{W_{n}(-\omega)  + W_{n}(0) \right\} +
  W^{*}_{m}( \omega)\left\{ W_{n}(0) + W^{*}_{n}(\omega) \right\} \right. \nonumber \\
  && \left. \hspace{0.5in} +
  W^*_{m}(0)\left\{W_{n}(-\omega) + W^{*}_{n}(\omega) \right\}
  \right]
 \nonumber \\
 & & \left. + W^{*}_{l}( \omega)
 \left[ W^{*}_{m}( \omega)
 \left\{W_{n}(0) +
 W^{*}_{n}(\omega)  \right\} + W_{m}(0)W_{n}(0)\right]  \right \}
  \label{Dlmnqeaibshort}
 \eeqa
\subsection{Electric-Field Induced Second Harmonic Generation}
\subsubsection{$\beta \leq 1$}
 \beqa
 D_{ll}^{IB}(-2\omega;\omega, \omega, 0) \hspace{-0.09 in}& =&\hspace{-0.09 in}
  \frac{2i\sqrt{\pi}}{\gamma_{lg}}
  \left\{ \frac{-i\Gamma_{lg}}{\omega(\omega + 2i\Gamma_{lg}) \gamma_{lg}}\left[
    W^{(2)^{*}}_{\beta_l}(\omega)
     + W^{(2)}_{\beta_l}(-\omega)
     \right] \right.
  \nonumber \\
& & \hspace{0.43 in}
 + \frac{2\Gamma_{lg}^2}{\omega^2(\omega + 2i\Gamma_{lg})^2} \left[
 W^{(1)}_{\beta_l}(-\omega) +
 W^{(1)^{*}}_{\beta_l}(\omega)
 - W^{(1)^{*}}_{\beta_l}(0)
  - W^{(1)}_{\beta_l}(0)
\right]  \nonumber
\\
 & & \left. \hspace{0.43 in}
 +  \frac{ 1}{{\omega^2}}
 \left[  W^{(1)}_{\beta_l}(-2\omega) + W^{(1)^{*}}_{\beta_l}(2\omega)
   - W^{(1)}_{\beta_l}(-\omega)  - W^{(1)^{*}}_{\beta_l}(\omega)
 \right]
 \right\}
  \label{Dllefshibbetashort}
 \eeqa
 \beqa
 D_{ln}^{IB}(-2\omega; \omega, \omega,0)  &=& \frac{-\pi}{\gamma_{lg}\gamma_{ng}}
 \left\{ \frac{1}{2\omega}
 \left[W^{(1)}_{\beta_n}(-\omega)
 \left \{  W^{(1)}_{\beta_l}(-2\omega)-  W^{(1)}_{\beta_l}(0)
 \right \}
 + W^{(1)^{*}}_{\beta_n}(\omega)
 \left \{ W^{(1)^{*}}_{\beta_l}(0) - W^{(1)^{*}}_{\beta_l}(2\omega)
 \right\}
 \right]
 \right.
 \nonumber
  \\
 & & \hspace{0.5in} + \frac{1}{\omega}
 \left[ \left\{
     W^{(1)}_{\beta_n}(-\omega)
  + W^{(1)}_{\beta_n}(0)\right\}
   \left\{
     W^{(1)}_{\beta_l}(- 2\omega)
   - W^{(1)}_{\beta_l}(-\omega)
 \right\} \right.
   \nonumber \\
 & &  \hspace{0.75 in}\left. +
 \left\{ W^{(1)^{*}}_{\beta_n}(\omega)
  + W^{(1)^{*}}_{\beta_n}(0) \right\}
     \left\{
    W^{(1)^{*}}_{\beta_l}(\omega)
  - W^{(1)^{*}}_{\beta_l}(2\omega)\right\}
\right]
  \nonumber \\
  & & \hspace{0.5in} + \frac{1}{2(\omega + i\Gamma_{ng})}
  \left[ \left\{ W^{(1)}_{\beta_l}(0)
  + W^{(1)^{*}}_{\beta_l}(0) \right\}
  \left\{ W^{(1)}_{\beta_n}(-\omega)
  - W^{(1)^{*}}_{\beta_n}(\omega)\right\}\right]
  \nonumber \\
  & & \hspace{0.5in} +\frac{1}{(\omega + 2i\Gamma_{ng})}\left[
  \left\{W^{(1)}_{\beta_n}(0) - W^{(1)^{*}}_{\beta_n}(0) +  %
  W^{(1)}_{\beta_n}(-\omega)
  - W^{(1)^{*}}_{\beta_n}(\omega) \right\}\times \right. \nonumber
  \\
  && \left. \left. \hspace{1.4in}
  \left\{ W^{(1)}_{\beta_l}(-\omega)
  + W^{(1)^{*}}_{\beta_l}(\omega) \right\}\right]
    \right \}
 \label{Dlnefshibbetashort}
 \eeqa
 \beqa
D_{lml}^{IB}(-2\omega; \omega, \omega,0)  &=&
\frac{-\pi}{\gamma_{lg}\gamma_{mg}}
 \left\{ \frac{1}{2\omega}
 \left[
       W^{(1)}_{\beta_m}(-\omega)
 \left \{   W^{(1)}_{\beta_l}(-2\omega) -  W^{(1)}_{\beta_l}(0)
 \right \}
 +   W^{(1)^{*}}_{\beta_m}(\omega)
 \left \{
      W^{(1)^{*}}_{\beta_l}(0)
    - W^{(1)^{*}}_{\beta_l}( 2\omega)
 \right\} \right] \right.
 \nonumber
  \\
 & & \hspace{0.5in} + \frac{1}{\omega}
 \left[ \left\{
     W^{(1)}_{\beta_m}(-2\omega)
  + W^{(1)}_{\beta_m}(-\omega)\right\}
   \left\{
     W^{(1)}_{\beta_l}(-2\omega)
   - W^{(1)}_{\beta_l}(-\omega)
 \right\} \right.
   \nonumber \\
 & &  \hspace{0.75 in}\left. +
 \left\{ W^{(1)^{*}}_{\beta_m}(2\omega)
  + W^{(1)^{*}}_{\beta_m}(\omega)  \right\}
     \left\{
    W^{(1)^{*}}_{\beta_l}(\omega)
  - W^{(1)^{*}}_{\beta_l}(2\omega)\right\}
\right]
  \nonumber \\
  & & \hspace{0.5 in}+ \frac{1}{2(\omega + i\Gamma_{lg})}
  \left[ \left\{
  W^{(1)}_{\beta_m}(-\omega)
  + W^{(1)^{*}}_{\beta_m}(\omega)
   \right\}
  \left\{
   W^{(1)}_{\beta_l}(-\omega)
  - W^{(1)^{*}}_{\beta_l}(\omega) \right\}\right]
  \nonumber \\
  & & \hspace{0.5 in}+\frac{1}{(\omega + 2i\Gamma_{lg})}\left[
  \left\{W^{(1)}_{\beta_m}(-2\omega)
   + W^{(1)^{*}}_{\beta_m}(\omega)
     \right\}
  \left\{ W^{(1)}_{\beta_l}(-\omega)
  - W^{(1)^{*}}_{\beta_l}(0) \right\} \right.
  \nonumber \\
  & & \hspace{1.4 in} \left.\left.
   + \left \{
     W^{(1)}_{\beta_m}(-\omega)
   + W^{(1)^{*}}_{\beta_m}(2\omega)
     \right\}
  \left\{
    W^{(1)}_{\beta_l}(0)
  - W^{(1)^{*}}_{\beta_l}(\omega)
   \right\}\right] \right \}
 \label{Dlmlefshibbetashort}
 \eeqa
 \beqa
D_{lmn}^{IB}(-2\omega; \omega, \omega, 0) &=&
 \frac{-i\pi^{3/2}}{\gamma_{lg}\gamma_{mg}\gamma_{ng}}
 \left\{
   W^{(1)}_{\beta_m}(-2\omega)
   W^{(1)}_{\beta_n}(-\omega)
 \left [
   W^{(1)}_{\beta_l}(-2\omega)
 + W^{(1)^{*}}_{\beta_l}(0) \right]
 \right.  \nonumber \\
 & & \hspace{0.75 in} +
  W^{(1)^{*}}_{\beta_l}(\omega)
  W^{(1)^{*}}_{\beta_m}(2\omega)
 \left [
  W^{(1)}_{\beta_n}(0)  +
  W^{(1)^{*}}_{\beta_n}(2\omega) \right ]
 \nonumber \\
 & & \hspace{0.75 in}
  + W^{(1)}_{\beta_m}(-\omega)
 \left[
 W^{(1)}_{\beta_l}(-2\omega) +
 W^{(1)^{*}}_{\beta_l}(\omega)
  \right]
 \left[
 W^{(1)}_{\beta_n}(-\omega) +
 W^{(1)}_{\beta_n}(0)
 \right]
   \nonumber \\
  & & \hspace{0.75 in}\left.
  + W^{(1)^{*}}_{\beta_m}(\omega)
 \left[
  W^{(1)}_{\beta_n}(-\omega) +
  W^{(1)^{*}}_{\beta_n}(2\omega)
  \right]
 \left[
  W^{(1)^{*}}_{\beta_l}(\omega)
  + W^{(1)^{*}}_{\beta_l}(0)\right]  \right \}
  \label{Dlmnefshibbetashort}
 \eeqa
\subsubsection{$\beta = 1$}
 \beqa
 D_{ll}^{IB}(-2\omega;\omega, \omega, 0)& =&
  \frac{2i\sqrt{\pi}}{\gamma_{lg}}
  \left\{ \frac{2\Gamma_{lg}}{\omega(\omega + 2i\Gamma_{lg}) \gamma_{lg}^2}\left[
    (\Omega^{*}_{lg} + \omega)W^{*}_{l}( \omega)
     + \frac{2i\gamma_{lg}}{\sqrt{\pi}}
  + (\Omega_{lg} - \omega)W_{l}( - \omega)
     \right] \right.
  \nonumber \\
& & \hspace{0.43 in}
 + \frac{2\Gamma_{lg}^2}{\omega^2(\omega + 2i\Gamma_{lg})^2} \left[
 W_{l}( - \omega) +
 W^{*}_{l}( \omega)
 - W^{*}_{l}(0)
  - W_{l}(0)
\right]  \nonumber
\\
 & & \hspace{0.43 in} \left.
 +  \frac{ 1}{{\omega^2}}
 \left[  W_{l}( - 2\omega) +
         W^{*}_{l}( 2\omega)
   - W_{l}( - \omega)
   - W^{*}_{l}( \omega)
 \right]
  \hspace{0.05 in} \right\}
  \label{Dllefshibshort}
 \eeqa
 \beqa
 D_{ln}^{IB}(-2\omega; \omega, \omega,0)  &=&  \frac{-\pi}{\gamma_{lg}\gamma_{ng}}
 \left\{ \frac{1}{2\omega}
 \left[
       W_{n}(-\omega)
 \left \{
       W_{l}( - 2\omega)
    -  W_{l}(0)
 \right \}
 +  W^{*}_{n}(\omega)
 \left \{
      W^{*}_{l}(0)
    - W^*_{l}(2\omega)
 \right\} \right] \right.
 \nonumber
  \\
 & &  + \frac{1}{\omega}
 \left[ \left\{
     W_{n}(-\omega)
  + W_{n}(0)\right\}
 \left\{
     W_{l}(- 2\omega)
   - W_{l}(- \omega)
 \right\}  +
 \left\{ W^{*}_{n}(\omega)
  + W^*_{n}(0) \right\}
     \left\{
    W^{*}_{l}( \omega)
  - W^{*}_{l}( 2\omega)\right\}
\right]
  \nonumber \\
  & & + \frac{1}{2(\omega + i\Gamma_{ng})}
  \left[ \left\{ W_{l}(0)
  + W^{*}_{l}(0) \right\}
  \left\{ W_{n}(-\omega)
  - W^{*}_{n}(\omega)\right\}\right]
  \nonumber \\
  & &  \left. +\frac{1}{(\omega + 2i\Gamma_{ng})}\left[
  \left\{W_{n}(0) - W^*_{n}(0) +
  W_{n}(-\omega) - W^{*}_{n}(\omega) \right\}
  \left\{ W_{l}(- \omega) + W^{*}_{l}( \omega) \right\}
  \right]
   \right \}
 \label{Dlnefshibshort}
 \eeqa
 \beqa
 D_{lml}^{IB}(-2\omega; \omega, \omega,0)\hspace{-0.09in}&=& \hspace{-0.09in}\frac{-\pi}{\gamma_{lg}\gamma_{mg}}
 \left\{ \frac{1}{2\omega}
 \left[ W_{m}(-\omega)
 \left \{ W_{l}( - 2\omega)-  W_{l}(0)
 \right \}
 +  W^{*}_{m}( \omega)
 \left \{  W^{*}_{l}(0) - W^*_{l}(2\omega)
 \right\} \right] \right.
 \nonumber
  \\
 & & \hspace{-0.09in}+ \frac{1}{\omega}
 \left[ \left\{
     W_{m}(- 2\omega)
  + W_{m}(- \omega)\right\}
  \left\{
     W_{l}(- 2\omega)
   - W_{l}(- \omega)
 \right\} +
 \left\{ W^{*}_{m}( 2\omega)
  + W^{*}_{m}( \omega)  \right\}
     \left\{
    W^{*}_{l}( \omega)
  - W^{*}_{l}( 2\omega)\right\}
\right]
  \nonumber \\
  & & \hspace{-0.09in}+ \frac{1}{2(\omega + i\Gamma_{lg})}
  \left[ \left\{
  W_{m}(- \omega)
  + W^{*}_{m}( \omega)
   \right\}
  \left\{
   W_{l}(- \omega)
  - W^{*}_{l}( \omega) \right\}\right]
  \nonumber \\
  & & \hspace{-0.09in} +\frac{1}{(\omega + 2i\Gamma_{lg})}\left[
  \left\{W_{m}(- 2\omega)
   + W^{*}_{m}( \omega)
     \right\}
  \left\{ W_{l}(- \omega)
  - W^*_{l}(0) \right\} \right.
  \nonumber \\
  & & \hspace{0.8 in} \left. \left.
   + \left \{
     W_{m}(- \omega)
   + W^{*}_{m}( 2\omega)
     \right\}
  \left\{
    W_{l}(0)
  - W^{*}_{l}( \omega)
   \right\}\right]
   \right \}
 \label{Dlmlefshibshort}
 \eeqa
 \beqa
 D_{lmn}^{IB}(-2\omega; \omega, \omega, 0) &=&
 \frac{-i\pi^{3/2}}{\gamma_{lg}\gamma_{mg}\gamma_{ng}}
 \left\{
   W_{m}(- 2\omega)
   W_{n}(-\omega)
 \left [
   W_{l}( - 2\omega)
 + W^*_{l}(0) \right]
 + W^{*}_{l}( \omega)W^{*}_{m}( 2\omega)
 \left [
  W_{n}(0)  +
  W^*_n(2\omega)
  \right] \right.
 \nonumber \\
 & & \hspace{0.01 in}
  + W_{m}(- \omega)
 \left[
 W_{l}( - 2\omega) +
 W^{*}_{l}( \omega)
  \right]
 \left[
 W_{n}(-\omega) +
 W_{n}(0)
 \right]
   \nonumber \\
  & & \left. \hspace{0.01 in}
  + W^{*}_{m}( \omega)
 \left[ W_{n}(-\omega) +  W^*_n(2\omega)
  \right]
 \left[
  W^{*}_{l}( \omega)+ W^{*}_{l}(0)\right]
   \right \}
  \label{Dlmnefshibshort}
 \eeqa
\subsection{Kerr Effect}
\subsubsection{Intensity dependent Refractive index}
\paragraph{$\beta \leq 1$}
 \beqa
 D_{ll}^{IB}(-\omega; \omega, -\omega, \omega) &=&
 \frac{i\sqrt{\pi}}{\gamma_{lg}}
 \left\{\frac{1}{\gamma_{lg}^2}
  W^{(3)}_{\beta_l} (- \omega)
  + \frac{1}{\gamma_{lg}^2}
  W^{(3)^{*}}_{\beta_{l}}( \omega) +
       \frac{\omega + 2i \Gamma_{lg}}{2i\omega\Gamma_{lg}}
 \left [\frac{1}{\gamma_{lg}}
 W^{(2)^{*}}_{\beta_l}(\omega)
 - \frac{1}{\gamma_{lg}}
 W^{(2)}_{\beta_l}(-\omega)
 \right ]\right.  \nonumber \\
 & & \hspace{0.3in} \left. +\frac{2 \Gamma_{lg}^2 - \omega^2 + 2i\Gamma_{lg}\omega}{4\Gamma_{lg}^2\omega^2}
 \left [  W^{(1)}_{\beta_l}(\omega)
 + W^{(1)^{*}}_{\beta_{l}}( - \omega) -
  W^{(1)^{*}}_{\beta_l}(\omega)
 - W^{(1)}_{\beta_l}(-\omega) \right]
  \right\}
 \label{Dllkeibbetashort}
 \eeqa
 \beqa
 D_{ln}^{IB}(-\omega;\omega,\omega,-\omega)
 &=&  \frac{-\pi}{\gamma_{lg}\gamma_{ng}}
 \left\{\left[ W^{(1)}_{\beta_n}(-\omega)
 +  W^{(1)}_{\beta_n}(\omega)\right]
 \left[ \frac{-1}{\gamma_{lg}}
 W^{(2)}_{\beta_l}(-\omega)
 \right]
  +
 \left[ W^{(1)^{*}}_{\beta_n}(\omega)
 +  W^{(1)^{*}}_{\beta_n}(-\omega)\right]
 \left[ \frac{-1}{\gamma_{lg}}
 W^{(2)^{*}}_{\beta_l}(\omega)
   \right] \right. \nonumber \\
 & &  +  \frac{1}{2i\Gamma_{ng}}\left[
 W^{(1)}_{\beta_n}(-\omega)
 +  W^{(1)}_{\beta_n}(\omega)
 -  W^{(1)^{*}}_{\beta_n}(-\omega)
 - W^{(1)^{*}}_{\beta_n}(\omega)
 \right]
 \left [  W^{(1)}_{\beta_l}(-\omega)
  +  W^{(1)^{*}}_{\beta_l}(\omega)
 \right]  \nonumber  \\
 & &  + \frac{1}{2(\omega + i\Gamma_{ng})}
 \left[ W^{(1)}_{\beta_n}(-\omega)
 - W^{(1)^{*}}_{\beta_n}(\omega) \right]
 \left[  W^{(1)}_{\beta_l}(\omega)
 + W^{(1)^{*}}_{\beta_l}(-\omega) \right]
   \nonumber \\
 & & \left. + \frac{1}{2\omega }\left[ W^{(1)}_{\beta_n}(-\omega)
 \left \{  W^{(1)}_{\beta_l}(-\omega)
 -  W^{(1)}_{\beta_l}(\omega) \right \}
 +
   W^{(1)^{*}}_{\beta_n}(\omega)
 \left \{  W^{(1)^{*}}_{\beta_l}(-\omega)
 -  W^{(1)^{*}}_{\beta_l}(\omega) \right \} \right ]
 \right\}
 \label{Dlnkeibbetashort}
 \eeqa
 \beqa
 D_{lml}^{IB}(-\omega;\omega,\omega,-\omega)
 &=&  \frac{-\pi}{\gamma_{lg}\gamma_{mg}}
 \left\{ \left[W^{(1)}_{\beta_m}(-2\omega) +  W^{(1)}_{\beta_m}(0)\right]
 \left[ \frac{-1}{\gamma_{lg}} W^{(2)}_{\beta_l}(-\omega) \right]
 +\left[ W^{(1)^{*}}_{\beta_m}(2\omega)
 +  W^{(1)^{*}}_{\beta_m}(0)\right]
 \left[ \frac{-1}{\gamma_{lg}}
 W^{(2)^{*}}_{\beta_l}(\omega)
 \right]   \right. \nonumber \\
 & &  + \frac{1}{2i\Gamma_{lg}}\left[
 \left\{ W^{(1)}_{\beta_m}(-2\omega)
 +  W^{(1)^{*}}_{\beta_m}(0) \right\}
 \left \{  W^{(1)}_{\beta_l}(-\omega)
 -  W^{(1)^{*}}_{\beta_l}(-\omega) \right \}
 \right.   \nonumber \\
 & &  \hspace{0.45 in} \left. +
 \left\{ W^{(1)^{*}}_{\beta_m}(2\omega)
 +  W^{(1)}_{\beta_m}(0)\right\}
 \left \{  W^{(1)}_{\beta_l}(\omega)
  -  W^{(1)^{*}}_{\beta_l}(\omega) \right \}
  \,\right ]   \nonumber  \\
 & & + \frac{1}{2(\omega + i\Gamma_{lg}) }
 \left[ W^{(1)}_{\beta_m}(0)
 + W^{(1)^{*}}_{\beta_m}(0) \right ]
 \left [  W^{(1)}_{\beta_l}(-\omega)
 -  W^{(1)^{*}}_{\beta_l}(\omega) \right ]
   \nonumber \\
 & &  \left. + \frac{1}{2\omega }\left[ W^{(1)}_{\beta_m}(0)
 \left \{  W^{(1)}_{\beta_l}(-\omega)
 -  W^{(1)}_{\beta_l}(\omega) \right \}
  + W^{(1)^{*}}_{\beta_m}(0)
 \left \{  W^{(1)^{*}}_{\beta_l}(-\omega)
 -  W^{(1)^{*}}_{\beta_l}(\omega) \right \} \right ]
 \right\}
 \label{Dlmlkeibbetashort}
 \eeqa
 \beqa
 D_{lmn}^{IB}(-\omega;\omega,\omega,-\omega)
 &=& \frac{-i\pi^{3/2}}{\gamma_{lg}\gamma_{mg}\gamma_{ng}}
 \left\{W^{(1)}_{\beta_m}(0)
 \left[
 W^{(1)}_{\beta_n}(-\omega)
 +  W^{(1)}_{\beta_n}(\omega)
 \right]
 \left [  W^{(1)}_{\beta_l}(-\omega)
  +  W^{(1)^{*}}_{\beta_l}(\omega)
 \right] \right. \nonumber  \\
 & & + W^{(1)^{*}}_{\beta_m}(0)
 \left[
 W^{(1)}_{\beta_n}(-\omega)
 +  W^{(1)^{*}}_{\beta_n}(\omega)
 \right]
 \left [  W^{(1)^{*}}_{\beta_l}(\omega)
  +  W^{(1)^{*}}_{\beta_l}(-\omega)
 \right]   \nonumber  \\
 & & + W^{(1)}_{\beta_m}(-2\omega)
 W^{(1)}_{\beta_n}(-\omega)
 \left[  W^{(1)}_{\beta_l}(-\omega)
 + W^{(1)^{*}}_{\beta_l}(-\omega) \right]
  \nonumber \\
 & & +  \left. W^{(1)^{*}}_{\beta_m}(2\omega)
 W^{(1)^{*}}_{\beta_l}(\omega)
 \left [  W^{(1)}_{\beta_n}(\omega)
 +  W^{(1)^{*}}_{\beta_n}(\omega)
 \right ] \right\}
 \label{Dlmnkeibbetashort}
 \eeqa
\paragraph{$\beta = 1$}
 \beqa
 D_{ll}^{IB}(-\omega;\omega ,-\omega,\omega) &=&
 \frac{i\sqrt{\pi}}{\gamma_{lg}}
 \left\{\frac{2(\Omega_{lg} - \omega)^2 -\gamma_{lg}^2}{\gamma_{lg}^4}
  W_{l}( - \omega)
  + \frac{2(\Omega^{*}_{lg} + \omega)^2 -\gamma_{lg}^2}{\gamma_{lg}^4}
  W^{*}_{l}( \omega) +
   \frac{4i \omega_{lg}}{\sqrt{\pi} \gamma_{lg}^3}  \right.  \nonumber \\
 & &  + \frac{2 \Gamma_{lg}^2 - \omega^2 + 2i\Gamma_{lg}\omega}{4\Gamma_{lg}^2\omega^2}
 \left [  W_{l}( \omega) + W^*_{l}(-\omega) -
  W^{*}_{l}( \omega) - W_{l}( - \omega) \right]
  \nonumber \\
 \hspace{-0.1in}& & +\left.
 \frac{\omega + 2i \Gamma_{lg}}{2i\omega\Gamma_{lg}}
 \left [\frac{2(\Omega_{lg} - \omega)}{\gamma_{lg}^2}W_{l}( - \omega)
 - \frac{2(\Omega^{*}_{lg}  + \omega)}{\gamma_{lg}^2}W^{*}_{l}( \omega)
 \right ] \right\}
 \label{Dllkeibshort}
 \eeqa
 \beqa
 D_{ln}^{IB}(-\omega;\omega,\omega,-\omega)
 &=&  \frac{-\pi}{\gamma_{lg}\gamma_{ng}}
 \left\{ \left[ W_{n}(-\omega)
 +  W_{n}(\omega)\right]
 \left[ \frac{2(\Omega_{lg} - \omega)}{\gamma^2_{lg}}
 W_{l}( - \omega)
 + \frac{2i}{\sqrt{\pi}\gamma_{lg}}
 \right] \right.
   \nonumber \\
 & & +
 \left[ W^{*}_{n}(\omega)
 +  W^{*}_{n}(-\omega)\right]
 \left[ \frac{2(\Omega^{*}_{lg} + \omega)}{\gamma^2_{lg}}
 W^{*}_{l}( \omega)
 + \frac{2i}{\sqrt{\pi}\gamma_{lg}}
  \right]  \nonumber \\
 & &  + \frac{1}{2i\Gamma_{ng}}\left[
 W_{n}(-\omega)
 +  W_{n}(\omega)
 -  W^{*}_{n}(-\omega)
 - W^{*}_{n}(\omega)
 \right]
 \left [  W_{l}(-\omega)
  +  W^{*}_{l}(\omega)
 \right]   \nonumber  \\
 & & + \frac{1}{2(\omega + i\Gamma_{ng})}
 \left[ W_{n}(-\omega)
 - W^{*}_{n}(\omega) \right]
 \left[  W_{l}(\omega)+ W^{*}_{l}(-\omega) \right]
  \nonumber \\
 & & + \left.  \frac{1}{2\omega }\left[ W_{n}(-\omega)
 \left \{  W_{l}(-\omega)
 -  W_{l}(\omega) \right \}
 +  W^{*}_{n}(\omega)
 \left \{  W^{*}_{l}(-\omega)
 -  W^{*}_{l}(\omega) \right \} \right ]
 \right\}
 \label{Dlnkeibshort}
 \eeqa
 \beqa
 D_{lml}^{IB}(-\omega;\omega,\omega,-\omega)
 \hspace{-0.09in}&=&\hspace{-0.09in} \frac{-\pi}{\gamma_{lg}\gamma_{mg}}
 \left\{\left[ W_{m}(-2\omega)
 +  W_{m}(0)\right]
 \left[ \frac{2(\Omega_{lg} - \omega)}{\gamma^2_{lg}}
 W_{l}( - \omega)
 + \frac{2i}{\sqrt{\pi}\gamma_{lg}}
 \right] \right.  \nonumber \\
 & &\hspace{-0.09in}  +
 \left[ W^{*}_{m}(2\omega)
 +  W^{*}_{m}(0)\right]
 \left[ \frac{2(\Omega^{*}_{lg} + \omega)}{\gamma^2_{lg}}
 W^{*}_{l}( \omega)
 + \frac{2i}{\sqrt{\pi}\gamma_{lg}}
   \right]   \nonumber \\
 & &\hspace{-0.09in}  + \frac{1}{2i\Gamma_{lg}}\left[
 \left\{ W_{m}(-2\omega)
 +  W^{*}_{m}(0)\right\}
 \left \{  W_{l}(-\omega)
 -  W^{*}_{l}(-\omega) \right \}
  +\left\{ W^{*}_{m}(2\omega)
 +  W_{m}(0)\right\}
 \left \{  W_{l}(\omega)
  -  W^{*}_{l}(\omega) \right \}
  \,\right ]   \nonumber  \\
 & &\hspace{-0.09in} + \frac{1}{2(\omega + i\Gamma_{lg}) }\left[ W_{m}(0)
 + W^{*}_{m}(0) \right ]
 \left [  W_{l}(-\omega)
 -  W^{*}_{l}(\omega) \right ]
   \nonumber \\
 & &\hspace{-0.09in}  \left. +\frac{1}{2\omega }\left[ W_{m}(0)
 \left \{  W_{l}(-\omega)-  W_{l}(\omega) \right \}
  + W^{*}_{m}(0)\left \{  W^{*}_{l}(-\omega)
  -  W^{*}_{l}(\omega) \right \}
  \right ]
 \right\}
 \label{Dlmlkeibshort}
 \eeqa
 \beqa
 D_{lmn}^{IB}(-\omega;\omega,\omega,-\omega)
 &=& \frac{-i\pi^{3/2}}{\gamma_{lg}\gamma_{mg}\gamma_{ng}} 
 \left\{W_{m}(0)
 \left[ W_{n}(-\omega) +  W_{n}(\omega)
 \right]
 \left [  W_{l}(-\omega)
  +  W^{*}_{l}(\omega)
 \right]  \right. \nonumber  \\
 & & + W^{*}_{m}(0)
 \left[
 W_{n}(-\omega)
 +  W^{*}_{n}(\omega)
 \right]
 \left [  W^{*}_{l}(\omega)
  +  W^{*}_{l}(-\omega)
 \right]   \nonumber  \\
 & & + \left. W_{m}(-2\omega)
 W_{n}(-\omega)
 \left[  W_{l}(-\omega)
 + W^{*}_{l}(-\omega) \right]
  + W^{*}_{m}(2\omega)
 W^{*}_{l}(\omega)
 \left [  W_{n}(\omega)
 +  W^{*}_{n}(\omega)
 \right ] \right\}
 \label{Dlmnkeibshort}
 \eeqa
\subsubsection{Pump-Probe}
\paragraph{$\beta \leq 1$}
  \beqa
D_{ll}^{IB}(-\omega_1;\omega_1 ,\omega_2,-\omega_2)  &=&
  \frac{2i\sqrt{\pi}}{ \gamma_{lg}}
 \left\{  \frac{2\,\omega_1}{(\omega_1 + \omega_2)(\omega_1 - \omega_2)}
  \left[
  \frac{1}{\gamma_{lg}}
  W^{(2)^{*}}_{\beta_l}(\omega_1)
   - \frac{1}{\gamma_{lg}}
  W^{(2)}_{\beta_l}(- \omega_1)
  \right] \right.
  \nonumber \\
 & & +\frac{2(\omega^4_1 - 2\,\omega^2_1\,\omega^2_2 + \omega^4_2 + 2i\Gamma^2_{lg}\omega^2_1 +
 6\Gamma^2_{lg}\omega^2_2 - 8i\Gamma_{lg}\omega_1\omega^2_2)}{(\omega_1 + \omega_2)^2(\omega_1 -
 \omega_2)^2 (\omega_1 + \omega_2 + 2i\Gamma_{lg})(\omega_1 - \omega_2 + 2i\Gamma_{lg})}  \left[
 W^{(1)}_{\beta_l}(-\omega_1)
 + W^{(1)^{*}}_{\beta_l}(\omega_1)\right]
  \nonumber \\
 & &  +\frac{(\omega^3_2 - \omega_1\omega^2_2  + 3i\Gamma_{lg}\omega^2_2
 - i\Gamma_{lg}\omega_1\omega_2 - 3\Gamma^2_{lg}\omega_2 +
 \Gamma_{lg}^2\omega_1)}{i\Gamma_{lg}\omega_2(\omega_1 - \omega_2)^2
 (\omega_1 + \omega_2 + 2i\Gamma_{lg})}
 \left[ W^{(1)}_{\beta_l}(-\omega_2) +
 W^{(1)^{*}}_{\beta_l}(\omega_2)\right]
  \nonumber \\
& &\left. +
 \frac{(\omega^3_2 + \omega_1\omega^2_2  - 3i\Gamma_{lg}\omega^2_2 - i\Gamma_{lg}\omega_1\omega_2
  - 3\Gamma^2_{lg}\omega_2 - \Gamma_{lg}^2\omega_1)}
 {i\Gamma_{lg}\omega_2(\omega_1 + \omega_2)^2 (\omega_1 - \omega_2 + 2i\Gamma_{lg})}
 \left[ W^{(1)}_{\beta_l}( \omega_2)
 + W^{(1)^{*}}_{\beta_l}( - \omega_2)\right]\right\}
  \label{Dllppibbetashort}
 \eeqa
 \beqa
 D_{ln}^{IB}(-\omega_1;\omega_1 ,\omega_2,-\omega_2)  &=&
  \frac{-\pi}{\gamma_{lg}\gamma_{ng}}
 \left\{ \left[\frac{-1}{\gamma_{lg}}
 W^{(2)}_{\beta_l}( - \omega_1)
  \right]
 \left[W^{(1)}_{\beta_n}(-  \omega_2)+ W^{(1)}_{\beta_n}( \omega_2)
 \right] \right. \nonumber \\
 && \hspace{0.5in} + \left[\frac{-1}{\gamma_{lg}}
 W^{(2)^{*}}_{\beta_l}( \omega_1)
  \right]
 \left[W^{(1)^{*}}_{\beta_n}( \omega_2)
 + W^{(1)^{*}}_{\beta_n}(-\omega_2)\right]
 \nonumber \\
 & &+ \hspace{0.0 in}\frac{1}{\omega_1 -\omega_2 + 2i\Gamma_{ng}}
 \left[W^{(1)}_{\beta_l}(-\omega_2)
 +  W^{(1)^{*}}_{\beta_l}(\omega_2) \right]
 \nonumber \\
 & &\hspace{0.6 in} \times
 \left[ W^{(1)}_{\beta_n}(- \omega_1) + W^{(1)}_{\beta_n}(\omega_2)
 - W^{(1)^{*}}_{\beta_n}( -\omega_2) -
 W^{(1)^{*}}_{\beta_n}( \omega_1)\right]
   \nonumber \\
 & & \hspace{0.00 in}+  \frac{1}{\omega_1 +\omega_2 + 2i\Gamma_{ng}}
 \left[W^{(1)}_{\beta_l}(\omega_2)
 +  W^{(1)^{*}}_{\beta_l}(-\omega_2) \right]
  \nonumber \\
 & & \hspace{0.6 in} \times
 \left[W^{(1)}_{\beta_n}(-\omega_2)
 + W^{(1)}_{\beta_n}(-\omega_1)
 - W^{(1)^{*}}_{\beta_n}( \omega_2)
 - W^{(1)^{*}}_{\beta_n}( \omega_1)\right]
  \nonumber \\
 & & \hspace{0.0 in}+ \frac{1}{ 2i\Gamma_{ng}}
 \left[W^{(1)}_{\beta_l}(-\omega_1)
 +  W^{(1)^{*}}_{\beta_l}(\omega_1) \right]
    \nonumber \\
 & & \hspace{.45 in} \times
 \left[W^{(1)}_{\beta_n}(\omega_2)
 + W^{(1)}_{\beta_n}(-\omega_2) -
 W^{(1)^{*}}_{\beta_n}( -\omega_2) -
 W^{(1)^{*}}_{\beta_n}( \omega_2)\right]
  \nonumber \\
 & & \hspace{0.09 in} + \frac{1}{\omega_1 -\omega_2}\left\{
 \left[ W^{(1)}_{\beta_l}(-\omega_1) - W^{(1)}_{\beta_l}(-\omega_2) \right]
  \left[  W^{(1)}_{\beta_n}(- \omega_1) + W^{(1)}_{\beta_n}(\omega_2) \right]   \right. \nonumber \\
 & & \hspace{.575 in} + \left.
 \left[W^{(1)^{*}}_{\beta_l}(\omega_2) -
 W^{(1)^{*}}_{\beta_l}( \omega_1)\right]
 \left[ W^{(1)^{*}}_{\beta_n}( \omega_1) +  W^{(1)^{*}}_{\beta_n}( -\omega_2) \right]
 \right\}
   \nonumber \\
 & & \hspace{0.09 in} + \frac{1}{\omega_1 + \omega_2}\left\{
 \left[W^{(1)}_{\beta_l}(-\omega_1) -
 W^{(1)}_{\beta_l}(\omega_2) \right]
 \left[ W^{(1)}_{\beta_n}(- \omega_1) +  W^{(1)}_{\beta_n}(-\omega_2) \right] \right.  \nonumber \\
 & & \hspace{.575 in} +  \left. \left.
 \left[W^{(1)^{*}}_{\beta_l}(-\omega_2) -
 W^{(1)^{*}}_{\beta_l}(\omega_1)\right] \left[
 W^{(1)^{*}}_{\beta_n}( \omega_1)
 -  W^{(1)^{*}}_{\beta_n}( \omega_2) \right]
  \right\} \right \}
 \label{Dlnppibbetashort}
 \eeqa
 \beqa
 D_{lml}^{IB}(-\omega_1;\omega_1 ,\omega_2,-\omega_2)  &=&
 \frac{-\pi}{\gamma_{lg}\gamma_{mg}}
 \left\{ \left[\frac{-1}{\gamma_{lg}}
 W^{(2)}_{\beta_l}( - \omega_1)
 \right]  \left[W^{(1)}_{\beta_m}(-\omega_1-\omega_2)
 + W^{(1)}_{\beta_m}(-\omega_1+\omega_2) \right] \right.
 \nonumber \\
 & &+ \left[\frac{-1}{\gamma_{lg}}
 W^{(2)^{*}}_{\beta_l}( \omega_1)
  \right] \left[W^{(1)^{*}}_{\beta_m}(\omega_1+\omega_2)
 + W^{(1)^{*}}_{\beta_m}(\omega_1-\omega_2) \right]
 \nonumber \\
 & &+ \frac{1}{\omega_1 -\omega_2 + 2i\Gamma_{lg}}
 \left\{ \left[W^{(1)^{*}}_{\beta_m}(0)
 + W^{(1)}_{\beta_m}(-\omega_1-\omega_2)\right]
 \left[W^{(1)}_{\beta_l}(-\omega_1)
 -  W^{(1)^{*}}_{\beta_l}(-\omega_2) \right] \right.
 \nonumber \\
 & &\hspace{0.9 in} + \left.
 \left[W^{(1)}_{\beta_m}(0) +
 W^{(1)^{*}}_{\beta_m}(\omega_1+\omega_2)\right]
 \left[W^{(1)}_{\beta_l}(\omega_2)
 -  W^{(1)^{*}}_{\beta_l}(\omega_1) \right]\right\}   \nonumber \\
 & & + \frac{1}{\omega_1 +\omega_2 + 2i\Gamma_{lg}}
 \left\{\left[W^{(1)^{*}}_{\beta_m}(0)
 + W^{(1)}_{\beta_m}(-\omega_1+\omega_2) \right]
 \left[W^{(1)}_{\beta_l}(-\omega_1)
 -  W^{(1)^{*}}_{\beta_l}(\omega_2) \right] \right.
  \nonumber \\
 & & \hspace{0.9 in} + \left.
 \left[W^{(1)}_{\beta_m}(0) +
 W^{(1)^{*}}_{\beta_m}(\omega_1-\omega_2)\right]
 \left[W^{(1)}_{\beta_l}(-\omega_2)
 - W^{(1)^{*}}_{\beta_l}(\omega_1) \right] \right\} \nonumber \\
 & &  +  \frac{1}{\omega_1 -\omega_2}\left\{
 \left[W^{(1)}_{\beta_m}(0) +
  W^{(1)}_{\beta_m}(-\omega_1-\omega_2) \right]
 \left[W^{(1)}_{\beta_l}(-\omega_1)
 - W^{(1)}_{\beta_l}(-\omega_2) \right] \right.   \nonumber \\
 & & \hspace{.425 in}+ \left.
 \left[W^{(1)^{*}}_{\beta_m}(0) +
 W^{(1)^{*}}_{\beta_m}(\omega_1+\omega_2)\right]
 \left[W^{(1)^{*}}_{\beta_l}(\omega_2)
 -  W^{(1)^{*}}_{\beta_l}(\omega_1) \right] \right\}
   \nonumber \\
 & &  + \frac{1}{ 2i\Gamma_{lg}}\left\{
 \left [W^{(1)^{*}}_{\beta_m}(\omega_1-\omega_2) +
 W^{(1)}_{\beta_m}(-\omega_1-\omega_2) \right]
 \left[W^{(1)}_{\beta_l}(-\omega_2)
 -  W^{(1)^{*}}_{\beta_l}(-\omega_2) \right]
  \right.   \nonumber \\
 & & \hspace{.275 in} + \left.
 \left[W^{(1)}_{\beta_m}(-\omega_1+\omega_2) +
 W^{(1)^{*}}_{\beta_m}(\omega_1+\omega_2)\right]
 \left[W^{(1)}_{\beta_l}(\omega_2)
 -  W^{(1)^{*}}_{\beta_l}(\omega_2) \right] \right\}
  \nonumber \\
 & &  + \frac{1}{\omega_1 + \omega_2}\left\{
 \left[W^{(1)}_{\beta_m}(0) +
 W^{(1)}_{\beta_m}(-\omega_1+\omega_2) \right]
 \left[W^{(1)}_{\beta_l}(-\omega_1)
 -  W^{(1)}_{\beta_l}(\omega_2) \right]
 \right.   \nonumber \\
 & & \hspace{.625 in} +\left. \left.
 \left[W^{(1)^{*}}_{\beta_m}(0) +
 W^{(1)^{*}}_{\beta_m}(\omega_1-\omega_2)\right]
 \left[W^{(1)^{*}}_{\beta_l}(-\omega_2)
 -  W^{(1)^{*}}_{\beta_l}(\omega_1) \right]
  \right\} \right \}
 \label{Dlmlppibbetashort}
 \eeqa
  \beqa
D_{lmn}^{IB}(-\omega_1;\omega_1 ,\omega_2,-\omega_2)  &=&
  \frac{-i\pi^{3/2}}{\gamma_{lg}\gamma_{mg}\gamma_{ng}} \left\{
  W^{(1)}_{\beta_m}(-\omega_1-\omega_2)
 \left[ W^{(1)}_{\beta_l}(-\omega_1) +
 W^{(1)^{*}}_{\beta_l}(-\omega_2)
  \right]
 \left[ W^{(1)}_{\beta_n}(- \omega_1)
 +  W^{(1)}_{\beta_n}(-\omega_2) \right]
 \right.
 \nonumber \\
 & &\hspace{0.0 in} + W^{(1)}_{\beta_m}(-\omega_1+\omega_2)
 \left[ W^{(1)}_{\beta_l}(-\omega_1) +
 W^{(1)^{*}}_{\beta_l}(\omega_2)\right]
 \left[
 W^{(1)}_{\beta_n}(- \omega_1)
 +  W^{(1)}_{\beta_n}(\omega_2) \right]
 \nonumber \\
 & &\hspace{0.0 in} + W^{(1)}_{\beta_m}(0)
 \left[ W^{(1)}_{\beta_l}(-\omega_1) +
 W^{(1)^{*}}_{\beta_l}(\omega_1)\right]
 \left[W^{(1)}_{\beta_n}(-\omega_2)
 +  W^{(1)}_{\beta_n}(\omega_2) \right] \nonumber \\
 & & \hspace{0.0 in} + W^{(1)^{*}}_{\beta_m}(\omega_1+\omega_2)
 \left[ W^{(1)^{*}}_{\beta_l}(\omega_1) +
 W^{(1)^{*}}_{\beta_l}(\omega_2)\right]
 \left[ W^{(1)}_{\beta_n}(\omega_2)
 +  W^{(1)^{*}}_{\beta_n}( \omega_1) \right] \nonumber \\
 & & \hspace{0.0 in} + W^{(1)^{*}}_{\beta_m}(\omega_1-\omega_2)
 \left[ W^{(1)^{*}}_{\beta_l}(\omega_1) +
 W^{(1)^{*}}_{\beta_l}(-\omega_2)\right]
 \left[ W^{(1)}_{\beta_n}(-\omega_2)
 +  W^{(1)^{*}}_{\beta_n}( \omega_1) \right] \nonumber \\
 & & \hspace{0.0 in} \left. + W^{(1)^*}_{\beta_m}(0)
 \left[ W^{(1)^{*}}_{\beta_l}(\omega_2) +
 W^{(1)^{*}}_{\beta_l}(-\omega_2)\right]
 \left[ W^{(1)}_{\beta_n}(-\omega_1)
 +  W^{(1)^{*}}_{\beta_n}( \omega_1) \right] \right\}
 \label{Dlmnppibbetashort}
 \eeqa
\paragraph{$\beta = 1$}
  \beqa
D_{ll}^{IB}(-\omega_1;\omega_1 ,\omega_2,-\omega_2)  &=&
  \frac{2i\sqrt{\pi}}{ \gamma_{lg}}
 \left\{  \frac{2\,\omega_1}{(\omega_1 + \omega_2)(\omega_1 - \omega_2)}
  \left[
  \frac{2(\Omega_{lg} - \omega_1)}{\gamma^2_{lg}}
  W_{l}( - \omega_1)
 - \frac{2(\Omega^{*}_{lg} + \omega_1)}{\gamma^2_{lg}}
  W^{*}_{l}( \omega_1)
 \right] \right.
  \nonumber \\
 & & +\frac{2(\omega^4_1 - 2\,\omega^2_1\,\omega^2_2 + \omega^4_2 + 2i\Gamma^2_{lg}\omega^2_1 +
 6\Gamma^2_{lg}\omega^2_2 - 8i\Gamma_{lg}\omega_1\omega^2_2)}{(\omega_1 + \omega_2)^2(\omega_1 -
 \omega_2)^2 (\omega_1 + \omega_2 + 2i\Gamma_{lg})(\omega_1 - \omega_2 + 2i\Gamma_{lg})} \left[
 W_{l}( - \omega_1)
 + W^{*}_{l}( \omega_1)\right]
  \nonumber \\
 & &  +\frac{(\omega^3_2 - \omega_1\omega^2_2  + 3i\Gamma_{lg}\omega^2_2
 - i\Gamma_{lg}\omega_1\omega_2 - 3\Gamma^2_{lg}\omega_2 +
 \Gamma_{lg}^2\omega_1)}{i\Gamma_{lg}\omega_2(\omega_1 - \omega_2)^2
 (\omega_1 + \omega_2 + 2i\Gamma_{lg})}
 \left[ W_{l}( - \omega_2) +
 W^{*}_{l}( \omega_2)\right]
  \nonumber \\
& &+\left.
 \frac{(\omega^3_2 + \omega_1\omega^2_2  - 3i\Gamma_{lg}\omega^2_2 - i\Gamma_{lg}\omega_1\omega_2
  - 3\Gamma^2_{lg}\omega_2 - \Gamma_{lg}^2\omega_1)}
 {i\Gamma_{lg}\omega_2(\omega_1 + \omega_2)^2 (\omega_1 - \omega_2 + 2i\Gamma_{lg})}
 \left[ W_{l}( \omega_2)
 + W^{*}_{l}(-\omega_2)\right]\right\}
  \label{Dllppibshort}
 \eeqa
 \beqa
D_{ln}^{IB}(-\omega_1;\omega_1 ,\omega_2,-\omega_2)  &=&
  \frac{-\pi}{\gamma_{lg}\gamma_{ng}}
 \left\{ \left[\frac{2(\Omega_{lg} - \omega_1)}{\gamma^2_{lg}}
 W_{l}( - \omega_1)
 + \frac{2i}{\sqrt{\pi}\gamma_{lg}}
 \right]
 \left[ W_{n}(-\omega_2)
 + W_{n}(\omega_2) \right]
 \right.
 \nonumber \\
 & & \hspace{0.0 in} +\left[\frac{2(\Omega^{*}_{lg} + \omega_1)}{\gamma^2_{lg}}
 W^{*}_{l}( \omega_1) + \frac{2i}{\sqrt{\pi}\gamma_{lg}}
 \right]
 \left[ W^*_{n}(\omega_2)
 + W^*_{n}(-\omega_2) \right]
 \nonumber \\
 & &\hspace{0.0 in} + \frac{1}{\omega_1 -\omega_2 + 2i\Gamma_{ng}}
 \left[W_{l}(-\omega_2) +  W^{*}_{l}( \omega_2) \right]
 \left[W_{n}(-\omega_1) +
 W_{n}(\omega_2)
 - W^*_{n}(-\omega_2) -
 W^*_{n}(\omega_1)\right]
   \nonumber \\
 & & \hspace{0.0in} + \frac{1}{\omega_1 +\omega_2 + 2i\Gamma_{ng}}
 \left[W_{l}(\omega_2) +  W^{*}_{l}(-\omega_2) \right]
 \left[W_{n}(-\omega_2)
 + W_{n}(-\omega_1)
 - W^*_{n}(\omega_2)
 - W^*_{n}(\omega_1)\right]
  \nonumber \\
 & & \hspace{0.0 in}+ \frac{1}{ 2i\Gamma_{ng}}
 \left[W_{l}( - \omega_1)
 +  W^{*}_{l}(  \omega_1) \right]
 \left[W_{n}(\omega_2)
 + W_{n}(-\omega_2)  -
 W^*_{n}(-\omega_2) -
 W^*_{n}(\omega_2)\right]
  \nonumber \\
 & & + \hspace{0.0 in} \frac{1}{\omega_1 -\omega_2}\left\{
 \left[W_{l}( - \omega_1) -
  W_{l}(  - \omega_2)  \right]
 \left[W_{n}(-\omega_1)
 + W_{n}(\omega_2) \right] \right.   \nonumber \\
 & & \hspace{.575 in}+ \left.
 \left[W^{*}_{l}( \omega_2) -
 W^{*}_{l}( \omega_1 )\right]
 \left[W^*_{n}(\omega_1)
 +  W^*_{n}(-\omega_2) \right] \right\}
   \nonumber \\
 & & \hspace{0.0 in} +\frac{1}{\omega_1 + \omega_2}\left\{
 \left[W_{l}( - \omega_1) -
 W_{l}( \omega_2)  \right]
 \left[W_{n}(-\omega_1)
 +  W_{n}(-\omega_2) \right]
 \right.  \nonumber \\
 & & \hspace{.575 in} + \left. \left.
 \left[W^{*}_{l}(-\omega_2) -
 W^{*}_{l}( \omega_1)\right]
 \left[W^*_{n}(\omega_1)
 -  W^*_{n}(\omega_2) \right]
  \right\} \right \}
 \label{Dlnppibshort}
 \eeqa
 \beqa
 D_{lml}^{IB}(-\omega_1;\omega_1 ,\omega_2,-\omega_2)  &=&
 \frac{-\pi}{\gamma_{lg}\gamma_{mg}}
\left\{ \left[\frac{2(\Omega_{lg} - \omega_1)}{\gamma^2_{lg}}
 W_{l}( - \omega_1)
 + \frac{2i}{\sqrt{\pi}\gamma_{lg}}
  \right]
 \left[W_{m}(- \omega_1  - \omega_2)
 + W_{m}(- \omega_1 +  \omega_2) \right] \right.
 \nonumber \\
 & &+ \left[\frac{2(\Omega^{*}_{lg} + \omega_1)}{\gamma^2_{lg}}
 W^{*}_{l}( \omega_1)
 + \frac{2i}{\sqrt{\pi}\gamma_{lg}}
  \right]
 \left[W^{*}_{m}( \omega_1  + \omega_2)
 + W^{*}_{m}( \omega_1 -  \omega_2) \right]
 \nonumber \\
 & & + \frac{1}{\omega_1 -\omega_2 + 2i\Gamma_{lg}}
 \left\{ \left[W^*_{m}(0)
 + W_{m}(- \omega_1 - \omega_2) \right]
 \left[W_{l}( - \omega_1)
 -  W^{*}_{l}(-\omega_2) \right] \right.
 \nonumber \\
 & &\hspace{0.8 in} + \left.
 \left[W_{m}(0) +
 W^{*}_{m}( \omega_1 + \omega_2)\right]
 \left[W_{l}( \omega_2)
 -  W^{*}_{l}( \omega_1) \right]\right\}  \nonumber \\
 & &  +  \frac{1}{\omega_1 +\omega_2 + 2i\Gamma_{lg}}
 \left\{\left[W^*_{m}(0)
 + W_{m}(- \omega_1 + \omega_2)  \right]
 \left[W_{l}( - \omega_1)
 -  W^{*}_{l}( \omega_2) \right] \right.
  \nonumber \\
 & & \hspace{0.8 in} + \left.
 \left[W_{m}(0) +
 W^{*}_{m}( \omega_1 - \omega_2) \right]
 \left[W_{l}( - \omega_2)
 - W^{*}_{l}( \omega_1) \right] \right\} \nonumber \\
 & &   + \frac{1}{\omega_1 -\omega_2}\left\{
 \left[W_{m}(0) +
  W_{m}(- \omega_1 - \omega_2)  \right]
 \left[W_{l}( - \omega_1)
 - W_{l}( - \omega_2) \right] \right.   \nonumber \\
 & & \hspace{.425 in}+ \left.
 \left[W^*_{m}(0) +
 W^{*}_{m}( \omega_1 + \omega_2)\right]
 \left[W^{*}_{l}( \omega_2)
 -  W^{*}_{l}( \omega_1) \right] \right\}
  \nonumber \\
 & & +  \frac{1}{ 2i\Gamma_{lg}}\left\{
 \left [W^{*}_{m}( \omega_1 - \omega_2)  +
 W_{m}(- \omega_1 - \omega_2)  \right]
 \left[W_{l}( - \omega_2)
 -  W^{*}_{l}(-\omega_2) \right]
  \right.   \nonumber \\
 & & \hspace{.475 in} + \left.
 \left[W_{m}(- \omega_1 + \omega_2) +
 W^{*}_{m}( \omega_1 + \omega_2)\right]
 \left[W_{l}( \omega_2)
 -  W^{*}_{l}( \omega_2) \right] \right\}
  \nonumber \\
 & &  + \frac{1}{\omega_1 + \omega_2}\left\{
 \left[W_{m}(0) +
 W_{m}(- \omega_1 + \omega_2)  \right]
 \left[W_{l}( - \omega_1)
 -  W_{l}( \omega_2) \right]
 \right.   \nonumber \\
 & & \hspace{.625 in} +\left. \left.
 \left[W^*_{m}(0) +
 W^{*}_{m}( \omega_1 - \omega_2)\right]
 \left[W^{*}_{l}(-\omega_2)
 -  W^{*}_{l}( \omega_1) \right]
  \right\} \right \}
 \label{Dlmlppibshort}
 \eeqa
  \beqa
D_{lmn}^{IB}(-\omega_1;\omega_1 ,\omega_2,-\omega_2)  &=&
 \frac{-i\pi^{3/2}}{\gamma_{lg}\gamma_{mg}\gamma_{ng}} \left\{
  W_{m}(- \omega_1 - \omega_2) \left[ W_{l}( - \omega_1) +
 W^{*}_{l}(-\omega_2)\right]
 \left[W_{n}(-\omega_1)
 +  W_{n}(-\omega_2) \right] \right.
 \nonumber \\
 & &\hspace{0.0 in} + W_{m}(- \omega_1 + \omega_2)
 \left[ W_{l}( - \omega_1) +
 W^{*}_{l}( \omega_2)\right]
 \left[W_{n}(-\omega_1)
 +  W_{n}(\omega_2) \right]
 \nonumber \\
 & &\hspace{0.0 in} + W_{m}(0)
 \left[ W_{l}( - \omega_1) +
 W^{*}_{l}( \omega_1)\right]
 \left[W_{n}(-\omega_2)
 +  W_{n}(\omega_2) \right] \nonumber \\
 & & \hspace{0.0 in} + W^{*}_{m}( \omega_1 + \omega_2)
 \left[ W^{*}_{l}( \omega_1) +
 W^{*}_{l}( \omega_2)\right]
 \left[ W_{n}(\omega_2)
 +  W^*_{n}(\omega_1) \right] \nonumber \\
 & & \hspace{0.0 in} + W^{*}_{m}( \omega_1 - \omega_2)
 \left[ W^{*}_{l}( \omega_1) +
 W^{*}_{l}(-\omega_2)\right]
 \left[ W_{n}(-\omega_2)
 +  W^*_{n}(\omega_1) \right] \nonumber \\
 & & \hspace{0.0 in} \left. + W^*_{m}(0)
 \left[ W^{*}_{l}( \omega_2) +
 W^{*}_{l}(-\omega_2)\right]
 \left[ W_{n}(-\omega_1)
 +  W^*_{n}(\omega_1) \right] \right\}
 \label{Dlmnppibshort}
 \eeqa
  \nopagebreak
 \twocolumn




\end{document}